\documentclass[english,pre]{revtex4-2}
\pdfoutput=1
\usepackage[T1]{fontenc}
\usepackage[latin1]{inputenc}
\usepackage{float}
\usepackage{bm}
\usepackage{amsmath}
\usepackage{amssymb}
\usepackage{stackrel}
\usepackage{graphicx}
\usepackage{wasysym}

\makeatletter
\usepackage{babel}

\makeatother

\usepackage{babel}
\begin{document}
\title{Driven dust-charge fluctuation and chaotic ion dynamics in the plasma
sheath and pre-sheath regions}
\author{Mridusmita Das, Suniti Changmai, and Madhurjya P.\ Bora}
\affiliation{Department of Physics, Gauhati University, Guwahati 781014, India}
\begin{abstract}
Possible existence of chaotic oscillations in ion dynamics in the
sheath and pre-sheath regions of a dusty plasma, induced by externally
driven dust-charge fluctuation, is presented in this work. In a complex
plasma, dust charge fluctuation occurs continuously with time due
to the variation of electron and ions current flowing into the dust
particles. In most of the works related to dust-charge fluctuation,
theoretically it is assumed that the average dust-charge fluctuation
follows the plasma perturbation, while in reality, the dust-charge
fluctuation is a semi-random phenomena, fluctuating about some average
value. The very cause of dust-charge fluctuation in a dusty plasma
also points to the fact that these fluctuations can be driven externally
by changing electron and ion currents to the dust particles. With
the help of a \emph{hybrid}-Particle in Cell-Monte Carlo (\emph{h}-PIC-MCC)
code in this work, we use the plasma sheath as a candidate for driving
the dust-charge fluctuation by periodically exposing the sheath-side
wall to UV radiation, causing photoemission of electrons, which in
turn drive the dust-charge fluctuation. We show that this \emph{driven}
dust-charge fluctuation can induce a chaotic response in the ion dynamics
in the sheath and the pre-sheath regions.
\end{abstract}
\maketitle

\section{Introduction}

Even after decades of active and fruitful research, complex plasmas
and plasma sheath continue to enjoy immense attention in the present-day
plasma physics research. In this work, we bring together the important
domains of complex plasmas (also known as dusty plasma), plasma sheath,
and the rich tapestry of nonlinear dynamics. While dusty plasma deals
with the physics of plasmas with relatively massive and charged dust
particles, electrons, ions, and neutrals, it can very well be studied
in the context of plasma sheath, as the presence of charged dust particles
significantly modifies the sheath properties and can unravel complex
plasma behavior in the sheath and pre-sheath regions \cite{basnet,das,mehdi,pandey,jana,Shukla,shukla1}.
Through a \emph{hybrid}-Particle in Cell-Monte Carlo Collision (\emph{h}-PIC-MCC)
code,\emph{ }we in this work use the plasma sheath as a candidate
to \emph{drive} the dust-charge fluctuation, which in turn induces
a chaotic response in the ion dynamics in the vicinity of the sheath.
The \emph{h-}PIC-MCC code in question has been developed by two of
the authors of this paper, which can handle dusty plasma dynamics
with various boundary conditions and has already been benchmarked
for different electron and ion dynamics in the electron-plasma, ion-acoustic,
and dust-ion-acoustic time scales \cite{suniti,suniti1,suniti2}.
As a nonlinear plasma environment is essentially multidimensional,
the possibility of chaos is invariably there. However, it is quite
difficult to directly observe chaotic oscillations in naturally occurring
plasmas as well as laboratory plasmas in contrast to carefully controlled
plasma environments with some kinds of driving mechanisms. Chaotic
oscillations are thus observed in various configurations such as in
plasma-diode experiments \cite{piel}, filamentary discharge plasmas
in presence of plasma bubble \cite{mariammal} etc. Theoretically,
there are numerous works which explore the possibility of chaos in
different plasma environments. Such examples can be found in chaotic
Alfvén waves in the context of driven Hamiltonian systems \cite{buti},
wave-wave interaction \cite{kuney}, quasilinear diffusions \cite{skiadas},
etc.

In reality, the amount of charge acquired by the dust particles in
a plasma is never constant, rather fluctuates continuously, owing
to the changing electron and ion currents to the dust particles. While,
the semi-random nature of dust-charge fluctuation in time is quite
natural and occurs due to the nonlinear nature of the plasma, theoretically
it has been customary to assume these fluctuations to be closely following
the plasma perturbation present in the system \cite{jana,varma,suniti1,suniti2,Shukla}.
As the amount charge on a dust particle varies according to the electron
and ion currents to the dust particle, one can also externally \emph{drive}
the fluctuation by varying these currents. One such situation is to
expose the dust particles to an intermittent (or periodic) bursts
of charged particles which can cause the dust-charge to fluctuate.
Due to the nonlinearity present in the system, there is a possibility
that this driven dust-charge fluctuation can induce a chaotic response
in the dynamics of the system. Though such a situation in the dust-acoustic
regime has been considered by Momeni et al.\ in 2007 \cite{momeni},
where they have shown a chaotic regime to exist in the oscillation
of the dust density, the subject has been largely unexplored. The
case of driven dust-charge fluctuation can also be compared to the
effect of charged debris moving in a plasma, usually relevant in space
plasmas, which has been a subject of some recent studies \cite{sen,filippo}.

In this work, we show that by creating a periodic bursts of high-energy
electrons through photoemission from a sheath-side wall, we can indeed
induce a chaotic response in the ion-dynamics which is localised to
the sheath and the pre-sheath regions. In Section II, we develop a
dusty plasma model for the plasma sheath, where we describe the sheath
structure and develop the sheath equations. In Section III, we consider
the case for driven dust-charge fluctuation and develop our chaotic
ion-dynamics model induced by the driven fluctuations. In Section
IV, we describe the \emph{h}-PIC-MCC simulation of the driven dust-charge
fluctuation and present the required results. Finally, in Section
V, we conclude.

\section{A dusty plasma model for plasma sheath}

Our plasma model consists of electrons, ions, and negatively-charged
dust particles. The characteristic timescale of interest is dust-ion-acoustic,
where the dust particles are involved in the plasma dynamics \emph{only}
through the Poisson equation and the dust-charge fluctuation equation,
due to their massive inertia. This is particularly true in our case,
as the dust density remains constant which is a reasonable approximation
in the dust-ion-acoustic time scale \cite{rinku,ruinku1}. The relevant
equations (in 1-D) are ion continuity and momentum equations, with
Boltzmannian electrons (owing to their negligible mass) 
\begin{eqnarray}
\frac{\partial n_{i}}{\partial t}+\frac{\partial}{\partial x}(n_{i}u_{i}) & = & 0,\\
\frac{\partial u_{i}}{\partial t}+u_{i}\frac{\partial u_{i}}{\partial x} & = & -\frac{1}{m_{i}n_{i}}\frac{\partial p_{i}}{\partial x}-\frac{e}{m_{i}}\frac{\partial\phi}{\partial x},\\
n_{e} & = & n_{0}e^{e\phi/T_{e}},
\end{eqnarray}
where the symbols have their usual meanings and the temperature is
expressed in energy unit. The ion equation of state is used as, 
\begin{equation}
p_{i}\propto n_{i}^{\gamma},
\end{equation}
where $\gamma$ is the ratio of specific heats. The final equation
of the model is the Poisson equation, 
\begin{equation}
\epsilon_{0}\frac{\partial^{2}\phi}{\partial x^{2}}=e(n_{e}-n_{i}+z_{d}n_{d}).
\end{equation}
The dust charge is $q_{d}=-ez_{d}$, where we have assumed that the
dust particles acquire a net negative charge. Note that the presence
of dust grains are incorporated into the model through the Poisson
equation only. We use a normalization where the densities are normalized
by their respective equilibrium values i.e.\ $n_{j}\to n_{j}/n_{j0}$,
where the subscript `0' refers to the equilibrium values and $j=e,i,d$
respectively for electrons, ions, and dust particles. The ion velocity
$u_{i}$ is normalised with the ion-sound velocity $c_{s}=\sqrt{T_{e}/m_{i}}$,
where $T_{i,e}$ are the ion and electron temperatures, measured in
the units of energy and are held constant. The length is normalised
with the electron Debye length and time is normalised with ion-plasma
frequency. The potential $\phi$ is normalised with $(T_{e}/e)$.
The dust-charge number $z_{d}$ is normalised with its equilibrium
value $z_{d0}=z_{d}|_{\phi=0}$. The normalized equations are now
\begin{eqnarray}
\frac{\partial n_{i}}{\partial t}+\frac{\partial}{\partial x}(n_{i}u_{i}) & = & 0,\label{eq:cont}\\
\frac{\partial u_{i}}{\partial t}+u_{i}\frac{\partial u_{i}}{\partial x}+\gamma\sigma n_{i}^{\gamma-2}\frac{\partial n_{i}}{\partial x} & = & -\frac{\partial\phi}{\partial x},\label{eq:mom}\\
n_{e} & = & e^{\phi},\\
\frac{\partial^{2}\phi}{\partial x^{2}} & = & n_{e}-\delta_{i}n_{i}+\delta_{d}z_{d},\label{eq:pois}
\end{eqnarray}
where $\delta_{i}=n_{i0}/n_{e0}$ and $\delta_{d}=n_{d}z_{d0}/n_{e0}$
are the ratios of equilibrium densities of ion and dust particles
to that of electrons. The quasi-neutrality condition is given as $\delta_{i}=1+\delta_{d}$.
Note that the dust density remains constant, while the dust-charge
fluctuates.

\begin{figure}
\begin{centering}
\includegraphics[width=0.49\textwidth]{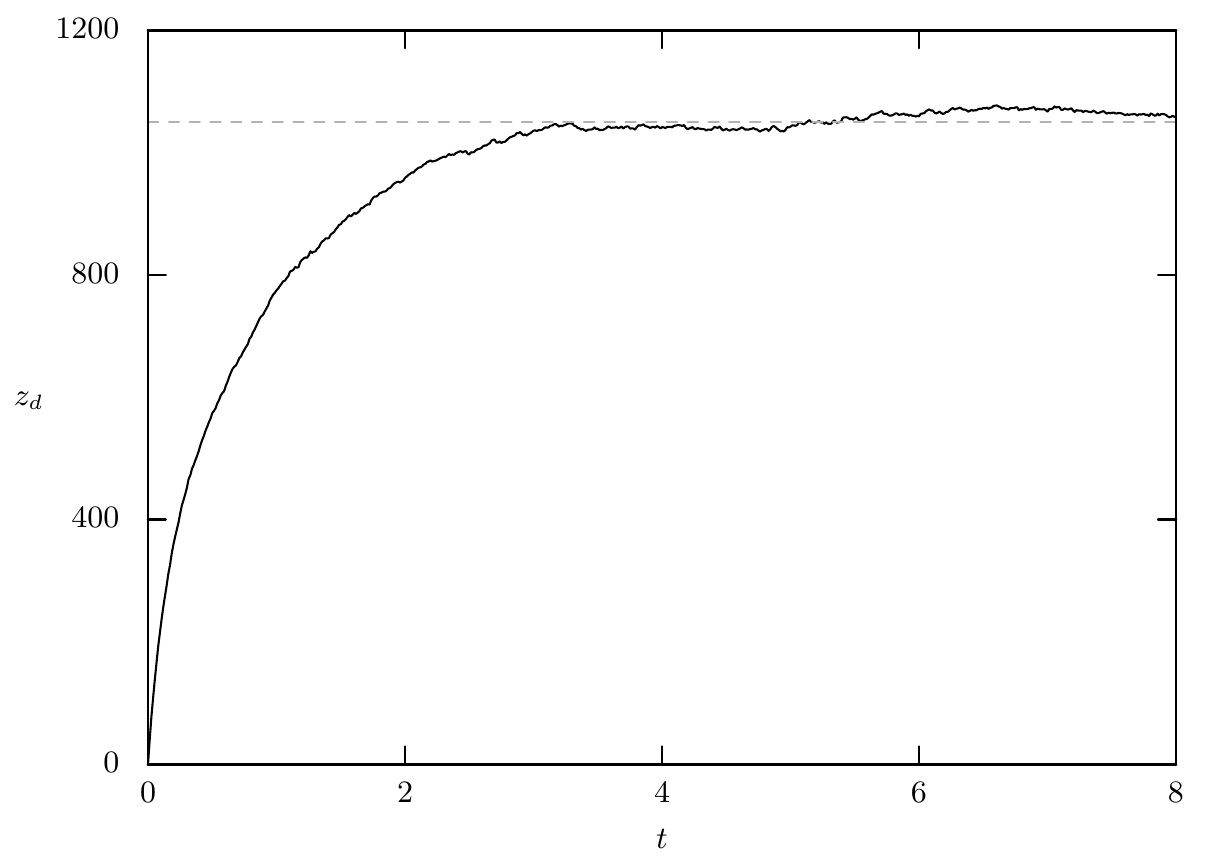} 
\par\end{centering}
\caption{\label{fig:Charging-of-dust}Charging of dust particles in a dusty
plasma.}
\end{figure}

The dust-charge $q_{d}$ for spherical dust particles can be expressed
in terms of the dust potential $\varphi_{d}$ 
\begin{equation}
q_{d}=C\,\Delta V=4\pi\epsilon_{0}r_{d}\varphi_{d},
\end{equation}
where $C$ is the grain capacitance and $\varphi_{d}=\phi_{g}-\phi$,
$\phi_{g}$ being the grain potential. We define the equilibrium dust-charge
number $z_{d0}$ in terms of the magnitude of the equilibrium dust
potential $\varphi_{d0}=\left|\varphi_{d}|_{\phi=0}\right|$, 
\begin{equation}
z_{d0}=4\pi\epsilon_{0}r_{d}e^{-1}\varphi_{d0},\label{eq:eqlb-charge}
\end{equation}
where $e$ is the magnitude of electronic charge and $r_{d}$ is the
radius of a dust-particle. By using the relation $q_{d}=-ez_{d}$,
we can normalize the expression for dust potential 
\begin{equation}
z_{d}=-\alpha\varphi_{d},
\end{equation}
with 
\begin{equation}
\alpha=\frac{4\pi\epsilon_{0}r_{d}T_{e}}{e^{2}z_{d0}}\sim N_{D},
\end{equation}
which approximately represents the total number of dust particles
$N_{D}$, inside a dust Debye sphere. We now consider the dust-charging
equation 
\begin{equation}
\frac{dq_{d}}{dt}=I_{e}+I_{i},\label{eq:dust-1}
\end{equation}
where $I_{e,i}$ are the electron and ion currents to the dust particles
which can be written as (dimensional) \cite{Shukla} 
\begin{eqnarray}
I_{i} & = & 4\pi r_{d}^{2}en_{i}\left(\frac{T_{i}}{2\pi m_{i}}\right)^{1/2}\left(1-\frac{e\varphi_{d}}{T_{i}}\right),\\
I_{e} & = & -4\pi r_{d}^{2}en_{e}\left(\frac{T_{e}}{2\pi m_{e}}\right)^{1/2}\exp\left(\frac{e\varphi_{d}}{T_{e}}\right).
\end{eqnarray}
Assuming Boltzmannian electron density $n_{e}=e^{\phi}$, the normalised
dust-charging equation Eq.(\ref{eq:dust-1}) can be written as 
\begin{equation}
\frac{d\varphi_{d}}{dt}=I_{e0}\left[\delta_{i}\delta_{m}\sigma^{1/2}n_{i}\left(1-\frac{\varphi_{d}}{\sigma}\right)-\exp(\phi+\varphi_{d})\right]\equiv f(n_{i},\phi,\varphi_{d}),\label{eq:charging}
\end{equation}
where 
\begin{equation}
I_{e0}=\frac{r_{d}e^{2}n_{e0}}{\epsilon_{o}T_{e}\omega_{pi}}\left(\frac{T_{e}}{2\pi m_{e}}\right)^{1/2}
\end{equation}
is the normalised equilibrium electron current to the dust particles,
$\delta_{m}=\sqrt{m_{e}/m_{i}}\approx0.023$, and $\sigma=T_{i}/T_{e}$.
A time evolution of average dust-charge $\overline{z}_{d}$ as obtained
from the \emph{hybrid}-PIC-MCC simulation (see Section III for a brief
description of the \emph{h-}PIC-MCC model) is shown in Fig.\ref{fig:Charging-of-dust}
\cite{suniti1}. In the code, both electrons and ions are considered
as thermal particles, distributed with their respective distributions.
The simulation parameters correspond to a typical laboratory situation
with plasma density ${n}_{{0}}\sim{10^{16}}\,{\rm m}^{-3}$, electron
and ion temperatures to be respectively $T_{e}\sim1\,{\rm eV}$ and
$T_{i}\sim0.01\,{\rm eV}$, and an $e$-$i$ mass ratio of $(m_{e}/m_{i})^{-1}\sim1835.16$.
This corresponds to the electron Debye length of the plasma as $\lambda_{{\rm D}e}\sim7.4\times10^{-5}\,{\rm m}$.
A typical run can have a simulation box length of $0.001-0.1\,{\rm m}$,
each species are represented with $10^{5}-10^{7}$ macro-particles
\cite{suniti,suniti1,suniti2} having particle weighting. The simulation
is carried out with equally spaced 1-D cells with a cell number of
$600-1000$. For example, in a simulation box length of $0.004\,{\rm m}$,
with $10^{5}$ macro particles for both electrons and ions, equally-spaced
$600$ cells, and the evolution time step of $\sim10^{-11}\,{\rm s}$,
we can have full spatial resolution with a temporal resolution of
the order of the electron time scale.

\subsection{Sheath equations and sheath structure}

Consider now a plasma sheath in steady state. Far away from the sheath,
the plasma potential vanishes and other plasma parameters approaches
their bulk (equilibrium) values i.e.\ $x\to\infty$, $\phi\to0$,
$u_{i}\to u_{0}\equiv M$, $p_{i}\to1$, $n_{i}\to1$, $z_{d}=z_{d}/z_{d0}\to1$.
$M$ is the Mach number which is the ratio of the ion velocity far
away from the sheath to that of ion-sound velocity. For a stationary
sheath, the steady state equations are, 
\begin{eqnarray}
\frac{\partial}{\partial x}(n_{i}u_{i}) & = & 0,\label{eq:cont-1}\\
u_{i}\frac{\partial u_{i}}{\partial x}+\gamma\sigma n_{i}^{\gamma-2}\frac{\partial n_{i}}{\partial x} & = & -\frac{\partial\phi}{\partial x},\label{eq:mom-1}\\
u_{i}\frac{\partial\varphi_{d}}{\partial x} & = & f(n_{i},\phi,\varphi_{d}).\label{eq:dust-charge}
\end{eqnarray}
From the continuity equation, we have 
\begin{equation}
n_{i}=M/u_{i}.\label{eq:ni}
\end{equation}
Integration of Eq.(\ref{eq:mom-1}) thus results the conservation
of total energy flux which is a combination of the kinetic flux, enthalpy
flux, and electrostatic flux, 
\begin{equation}
\phi=\frac{1}{2n_{i}^{2}}M^{2}\left(n_{i}^{2}-1\right)+\frac{\gamma\sigma}{(\gamma-1)}\left(1-n_{i}^{\gamma-1}\right).\label{eq:ni-1}
\end{equation}
An expression for $n_{i}$ as a function of $\phi$ can be found from
Eqs.(\ref{eq:mom-1},\ref{eq:ni}), $n_{i}={\cal F}(\phi)$. For arbitrary
$\gamma$, the above equation has to be solved numetically. For $\gamma=3$
however, we can find an analytical expression for $n_{i}(\phi)$ as,
\begin{equation}
n_{i}=\frac{1}{2\sqrt{3\sigma}}\left[\left\{ \left(M+\sqrt{3\sigma}\right)^{2}-2\phi\right\} ^{1/2}-\left\{ \left(M-\sqrt{3\sigma}\right)^{2}-2\phi\right\} ^{1/2}\right].
\end{equation}
The signs in front of the square roots are fixed through the boundary
condition on $n_{i}$. As the ion density can be expressed as a function
of the plasma potential, $n_{i}\equiv n_{i}(\phi)$, Poisson's equation
can be integrated to get, 
\begin{equation}
\frac{1}{2}\left(\frac{d\phi}{dx}\right)^{2}+V(\phi,M,\sigma,\gamma)=0,
\end{equation}
where $V(\phi,M,\sigma,\gamma)$ is the equivalent Sagdeev potential
or pseudo potential for a sheath, given by, 
\begin{equation}
V(\phi,\varphi_{d},M,\sigma,\gamma)=1-e^{\phi}+\delta_{i}\int_{0}^{\phi}n_{i}(\phi)\,d\phi-\delta_{d}\int_{0}^{\phi}z_{d}(\phi)\,d\phi.\label{eq:sagdeev}
\end{equation}
For real solution, we must have 
\begin{equation}
V(\phi,M,\sigma,\gamma)<0\label{eq:sag}
\end{equation}
for all values of $\phi$. We can also determine the minimum velocity
for the ions ($u_{0}\equiv M$) at the sheath boundary (the Bohm condition)
from this condition. The boundary condition on $V$ is: at $\phi=0$,
$V(\phi)=0$.

A few noteworthy points are in order at this moment. If the dust-charge
fluctuation is absent, it means the electron and ion currents to the
dust particles always balance each other so that at all time, we have
\begin{equation}
f(n_{i},\phi,\varphi_{d})=0.
\end{equation}
This equation can be numerically solved for $z_{d}$ (or for $\varphi_{d}$)
as a functions of $\phi$ and the Sagdeev potential can be constructed
numerically. However, in presence of dust-charge fluctuation, the
problem \emph{has} to be solved numerically. Multiplying Eq.(\ref{eq:dust-charge})
with $n_{i}$ and using Eq.(\ref{eq:cont-1}), we can write 
\begin{equation}
\frac{\partial}{\partial x}(n_{i}u_{i}\varphi_{d})=n_{i}f(n_{i},\phi,\varphi_{d}),
\end{equation}
where we note that $M=n_{i}u_{i}$. Thus, using Poisson equation,
Eq.(\ref{eq:sagdeev}), and the above equation, one can summarily
construct the following numerical model
\begin{eqnarray}
\frac{d^{2}\phi}{dx^{2}} & = & e^{\phi}-\delta_{i}n_{i}-\alpha\delta_{d}\varphi_{d},\\
\frac{d\varphi_{d}}{dx} & = & \frac{n_{i}}{M}f,\label{eq:ini}\\
n_{i} & = & {\cal F}(\phi).\label{eq:alg}
\end{eqnarray}
This problem is a coupled boundary and initial value problem involving
a nonlinear Poisson equation, which needs to be solved with a hybrid
approach. We have solved this model with a finite-difference algorithm
with a Newton-iteration for the nonlinear Poisson equation with Dirichlet
boundary conditions. During every Newton iteration of the boundary
value problem, we use a single, 4th order Runge-Kutta step for the
initial value equation, Eq.(\ref{eq:ini}). Of course, in every step,
the nonlinear algebraic equation, Eq.(\ref{eq:alg}) needs to be solved
for $n_{i}$, which we have solved using a nonlinear solver. However,
as we shall see, although the boundary values for the Poisson equation
can be estimated quite accurately, the same can not be said for the
initial value for Eq.(\ref{eq:ini}). This has to be estimated iteratively
following the negativity condition (\ref{eq:sag}) for a numerically
constructed Sagdeev potential.

\begin{figure}
\includegraphics[width=0.49\textwidth]{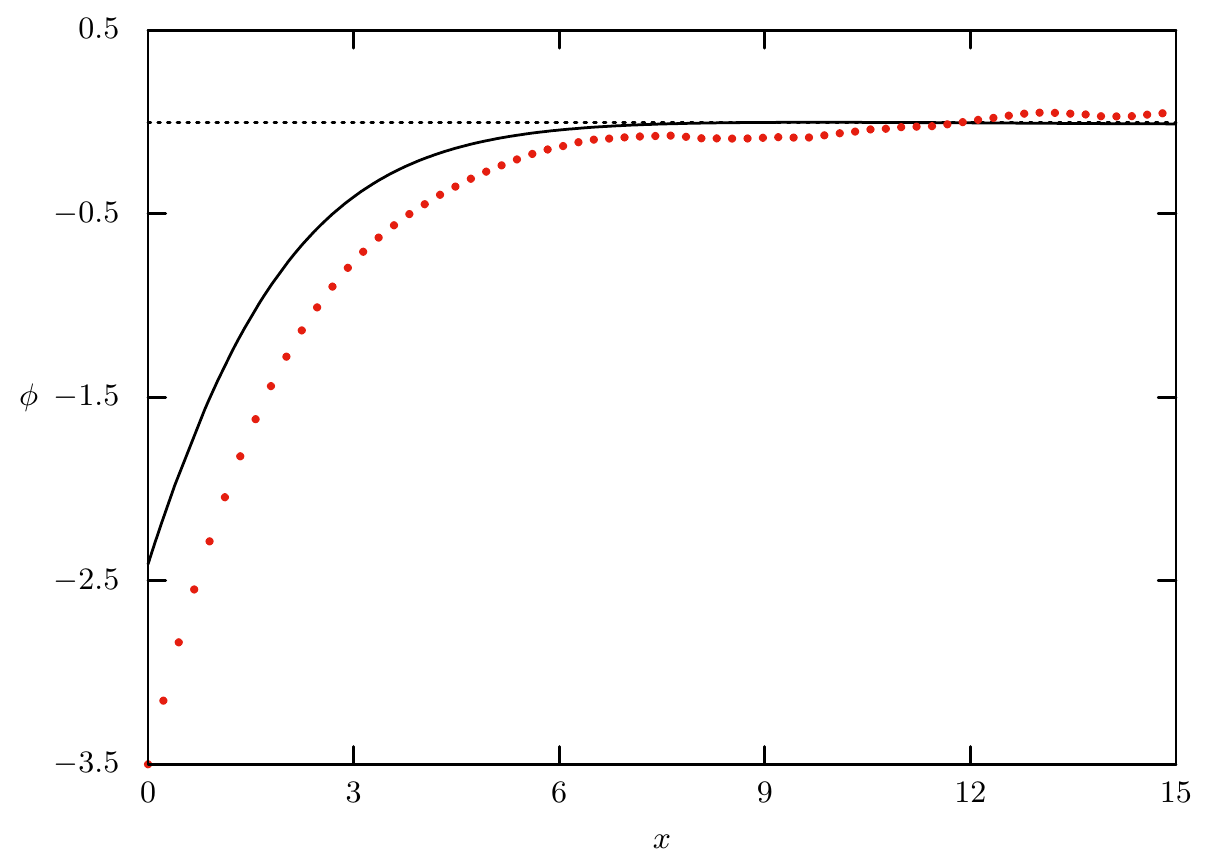}\hfill{}\includegraphics[width=0.49\textwidth]{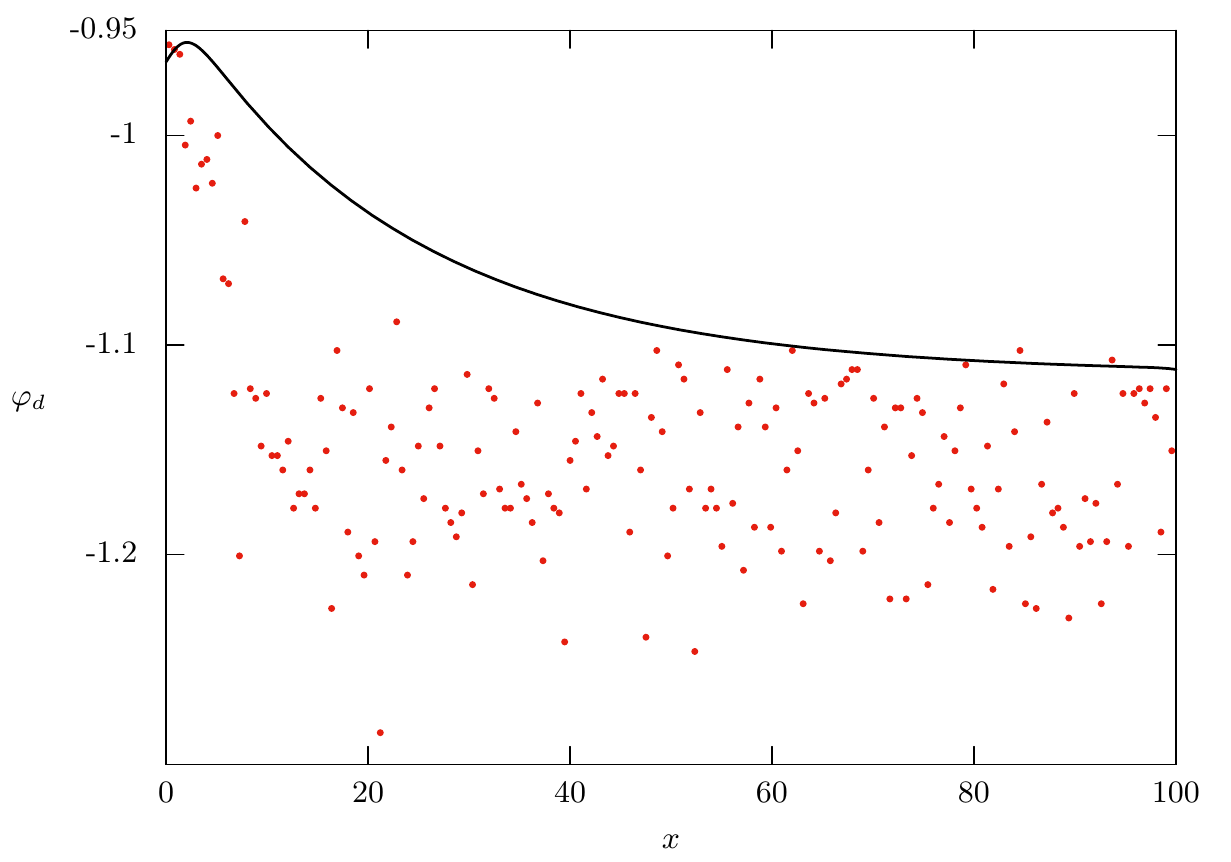}

\caption{\label{fig:Plasma-potential-}Plasma potential $\phi$ and dust potential
$\varphi_{d}$ with dust-charge fluctuation.}
\end{figure}

The boundary values for the plasma potential can be estimated by considering
the current at the wall and infinity. Far away from the sheath, the
plasma potential \emph{must} approach the bulk potential i.e.\ zero,
$\phi_{x\to\infty}=0$. Assuming the current at the wall be zero,
for a stationary sheath we have
\begin{equation}
j_{e}+j_{i}+j_{d}=0,
\end{equation}
where $j_{e,i,d}$ are the electron, ion, and dust currents to the
wall. However in view of inertia of the massive dust particles in
comparison to the electrons and ions, it can be safely assumed that
in the electron and ion timescale, the contribution to the wall current
by the dust particles is negligibly small,
\begin{equation}
j_{e}+j_{i}\approx0\label{eq:neu}
\end{equation}
for all practical purposes. The electrons, which reach the wall with
a minimum velocity $v_{\textrm{min}}$ by overcoming the negative
potential at the wall $\phi_{w}$, contribute to electron-current
at the wall. So, we have 
\begin{equation}
v_{\textrm{min}}=\left(-\frac{2e\phi_{w}}{m_{e}}\right)^{1/2},
\end{equation}
so that we have for the electron current 
\begin{equation}
j_{e}=-e\int_{v_{\textrm{min}}}^{\infty}\int_{-\infty}^{\infty}\int_{-\infty}^{\infty}vf_{e}(v)\,d\bm{v},
\end{equation}
where $f_{e}(v)$ is the electron velocity distribution function.
For a Maxwellian velocity distribution, we have the expression 
\begin{equation}
j_{e}=-n_{e0}e\left(\frac{T_{e}}{2\pi m_{e}}\right)^{1/2}\exp\left(\frac{e\phi_{w}}{T_{e}}\right).
\end{equation}
The ion current at the wall is given by 
\begin{equation}
j_{i}=en_{i0}u_{i}\left(\frac{T_{e}}{m_{i}}\right)^{1/2}=en_{0}u_{0}\left(\frac{T_{e}}{m_{i}}\right)^{1/2},
\end{equation}
where $u_{0}$ is the ion velocity (the Mach number) at the sheath
boundary. From the neutrality condition (\ref{eq:neu}), we can solve
for the wall plasma potential (normalised) as 
\begin{equation}
\phi_{w}=-2.84+\ln M.
\end{equation}

As the plasma potential vanishes far away from the sheath, we can
determine the dust potential as well, from the charging equation as
at $\infty$, $\partial\varphi_{d}/\partial x\to0$, so that $f(\varphi_{d})|_{x\to\infty}=0$.
This determines the $\varphi_{d}|_{x\to\infty}=\varphi_{d\infty}$
as 
\begin{equation}
\varphi_{d\infty}=\sigma-W(z),
\end{equation}
where $W(z)$ is the Lambert $W$ function with 
\begin{equation}
z=\frac{\sigma^{1/2}}{\delta_{i}\delta_{m}n_{i}}e^{\sigma}.
\end{equation}
Numerically however it is \emph{not} possible to use this value as
an initial value for solving the dust-charging equation as we need
to start at very large distance from the wall. We realize that although
the dust current at the wall is negligible, the dust potential $\varphi_{dw}$
at the wall need not be so and is expected to be small negative. So,
we continue solving the whole model starting with $\varphi_{dw}=0$
iteratively with the condition that the Sagdeev potential $V(\phi,\varphi_{d},M,\sigma,\gamma)<0$
\cite{Sagdeev} in the entire sheath region for the defined parameters.
The results of this calculation is shown in Fig.\ref{fig:Plasma-potential-}
where we plot the plasma potential $\phi$ and the dust potential
$\varphi_{d}$ as we go away from the sheath to the bulk plasma. The
solid lines in each panel indicate the theoretical curves obtained
for $\gamma=5/3$ for $M=1.5$ and the dots indicate the equivalent
results from the \emph{h-}PIC-MCC simulation with parameters as mentioned
before. We must \emph{note} that in the simulation, there is no fixed
Mach number for the ions as they enter the sheath with different Mach
numbers starting with the minimum.

\begin{figure}
\begin{centering}
\includegraphics[width=0.49\textwidth]{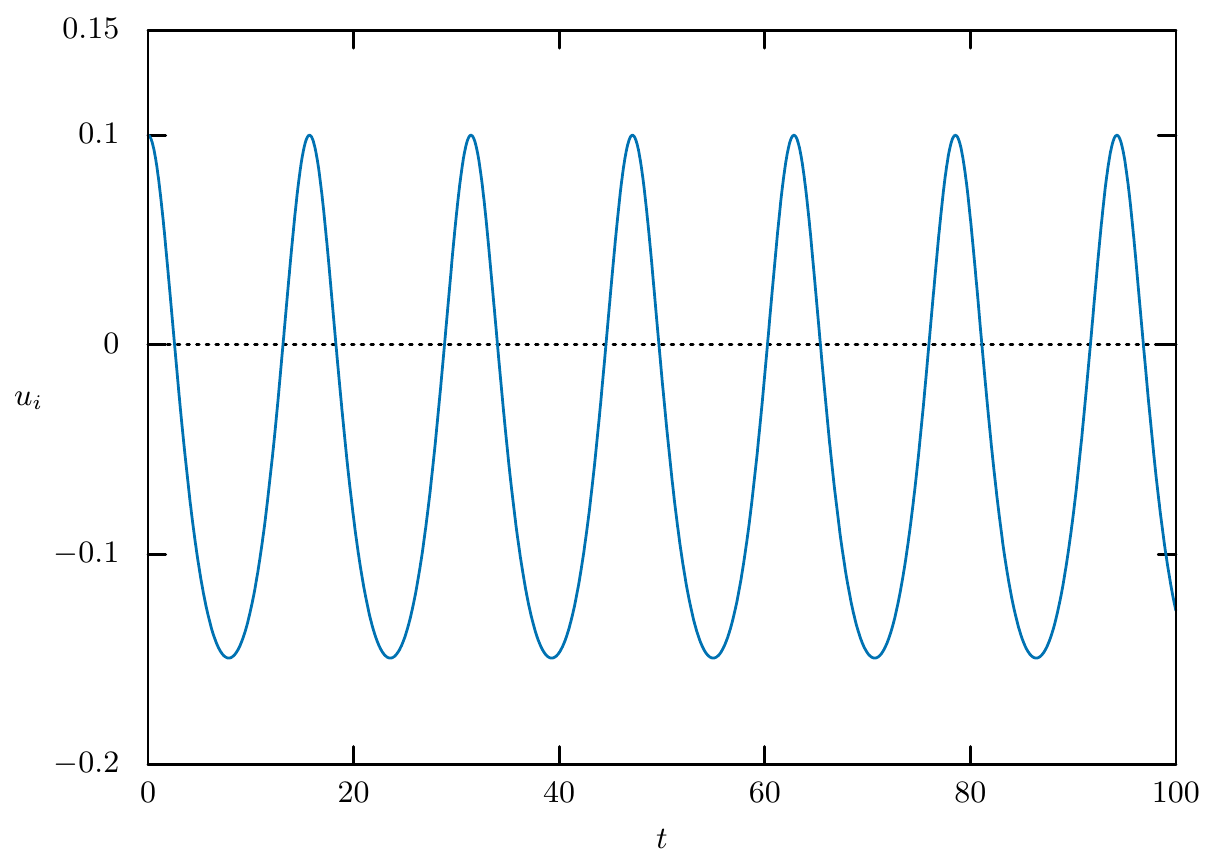}\hfill{}\includegraphics[width=0.49\textwidth]{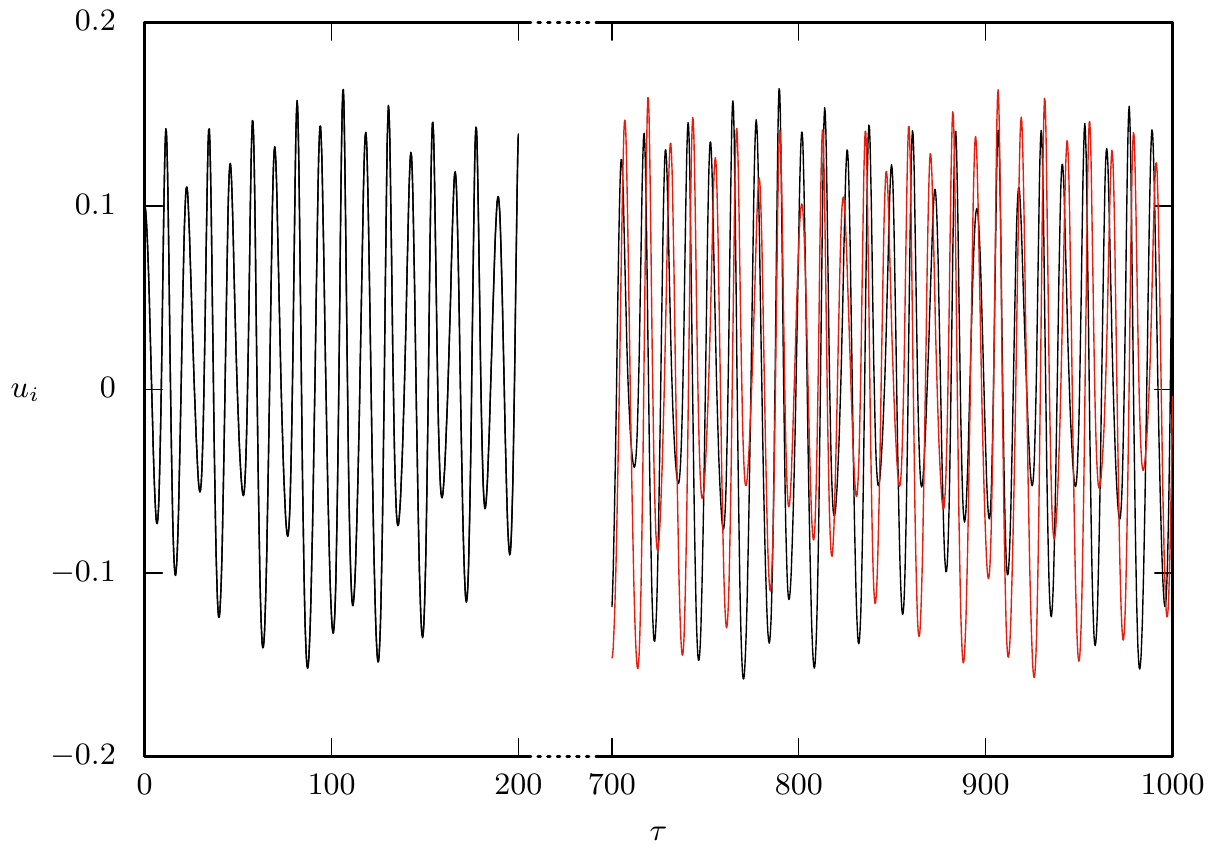} 
\par\end{centering}
\caption{\label{fig:The-periodic-oscillations}The periodic oscillations of
the system represented by Eq.(\ref{eq:chaos}) in absence of dust-charge
fluctuations (left). On the right, the same oscillations with driven
dust-charge fluctuation is presented where the sensitivity to the
initial conditions can be clearly seen. All plasma parameters are
as in the case of \emph{h-}PIC-MCC simulation.}
\end{figure}

\section{Driven dust-charge fluctuation and chaos}

A dynamical model demonstrating chaos in dusty plasma driven by dust-charge
fluctuation was formulated by Momeni et al.\ in 2007 \cite{momeni,dipak}.
In that work, they reduced the dusty plasma dynamical model to a single
second order, non-autonomous differential equation in dust density,
which exhibits chaotic dynamics in the dust-acoustic regime. This
single differential equation is described as a Van der Pol-Mathieu
(VdPM) equation owing to its Van der Pol-like term and a Mathieu-like
non-autonomous term 
\begin{equation}
\frac{d^{2}n_{d}}{dt^{2}}-\left(\alpha-\beta n_{d}^{2}\right)\frac{dn_{d}}{dt}+\omega_{0}^{2}(1+h\,\cos\gamma t)n_{d}=0,
\end{equation}
where $n_{d}$ is the dust density, $\omega_{0}\sim\omega_{pd}$ the
characteristic oscillation frequency of the system of the order of
dust-plasma frequency $\omega_{pd}$. The non-autonomous term results
from the time-dependent dust-charge fluctuation term, represented
by $h\cos\gamma t$, with $h$ as the amplitude of the fluctuation
and $\gamma$ as the fluctuation frequency. As seen from the equation,
the dust-charge fluctuation term is supposed to oscillate harmonically.
The authors assumed that the dust particles are continuously created
and destroyed through a source term $\alpha n_{d}$ and a loss term
$\beta n_{d}^{3}/3$ in the 1-D dust continuity equation \cite{momeni}
\begin{equation}
\frac{\partial n_{d}}{\partial t}+\frac{\partial}{\partial x}(n_{d}u_{d})=\alpha n_{d}-\frac{1}{3}\beta n_{d}^{3},
\end{equation}
where the source terms is assumed to appear due to the production
of charged dust grains through electron absorption and the loss terms
is due to a three-body recombination term.

It is important to realize that in the above work, though the authors
\emph{do not} describe why the time-dependent dust-charge fluctuation
term is harmonic, it can be argued to be originating from an externally
driven source -- such as \emph{photoemission}, which is what we are
investigating in this work. In this work, we provide a prescription
where chaotic oscillations are experimentally realizable. Our parameters
are however in the ion-acoustic regime unlike the above model. In
what follows, we construct a model which can demonstrate chaotic ion
dynamics through \emph{driven} dust-charge fluctuation. There is however
an important difference between the above mentioned work and ours
is that here we are using the dust-charge fluctuation as a mechanism
to drive the chaotic dynamics in ion velocity whereas in the former
the same mechanism is to drive the dynamics is dust density.

\begin{figure}
\begin{centering}
\includegraphics[width=0.49\textwidth]{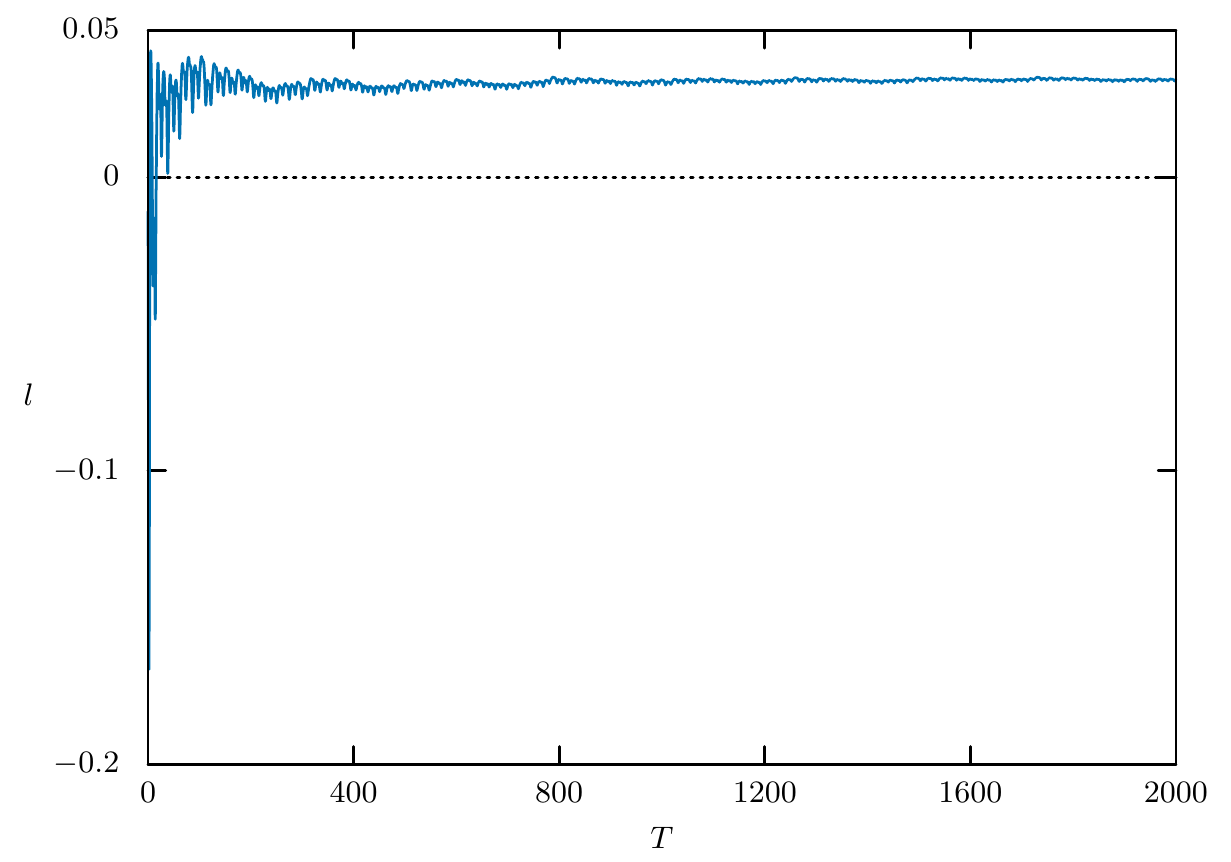}\hfill{}\includegraphics[width=0.49\textwidth]{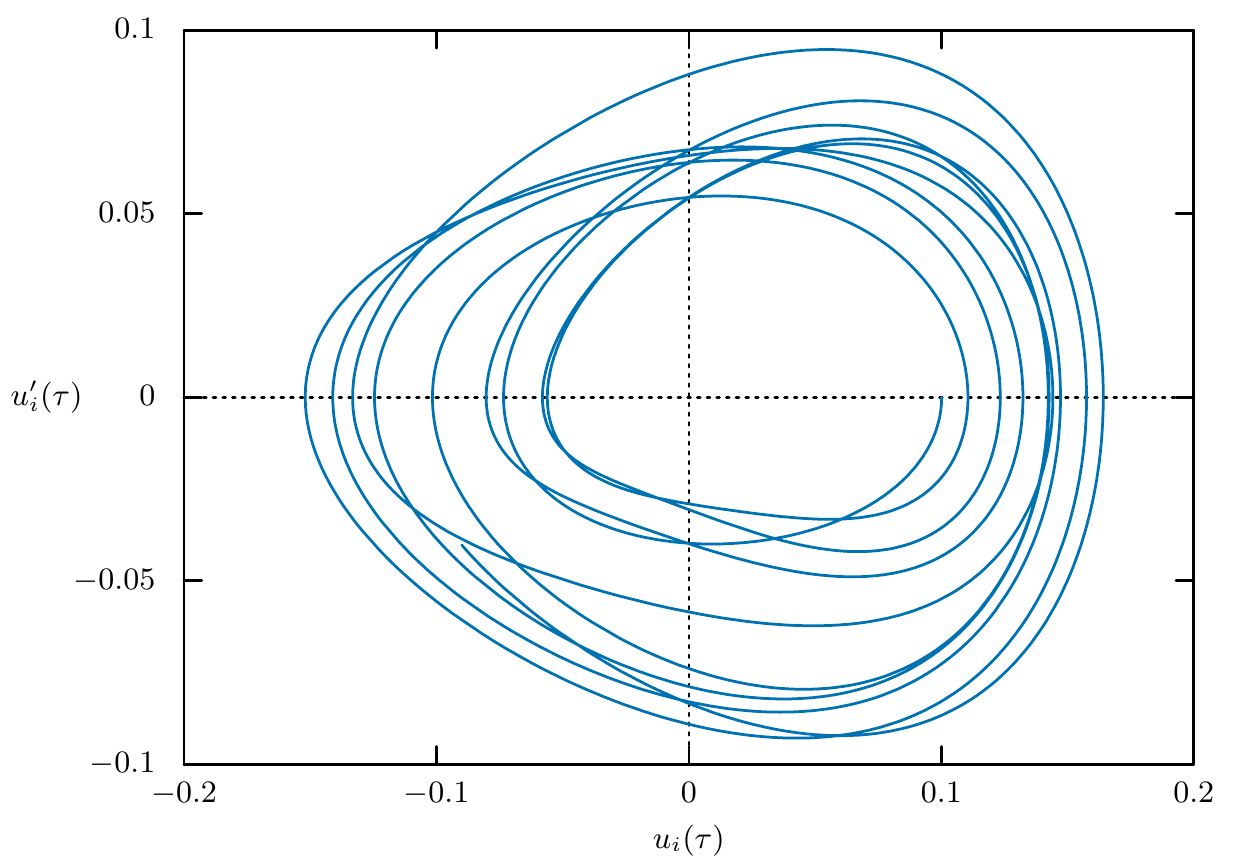} 
\par\end{centering}
\caption{\label{fig:The-maximal-Lyapunov}The maximal Lyapunov exponent $l$
of the system (left) shown for the first $\tau\sim500$ and the phase
portrait in the chaotic regime (right). The value of $l\sim0.033$.}
\end{figure}

\begin{figure}
\begin{centering}
\includegraphics[width=0.49\textwidth]{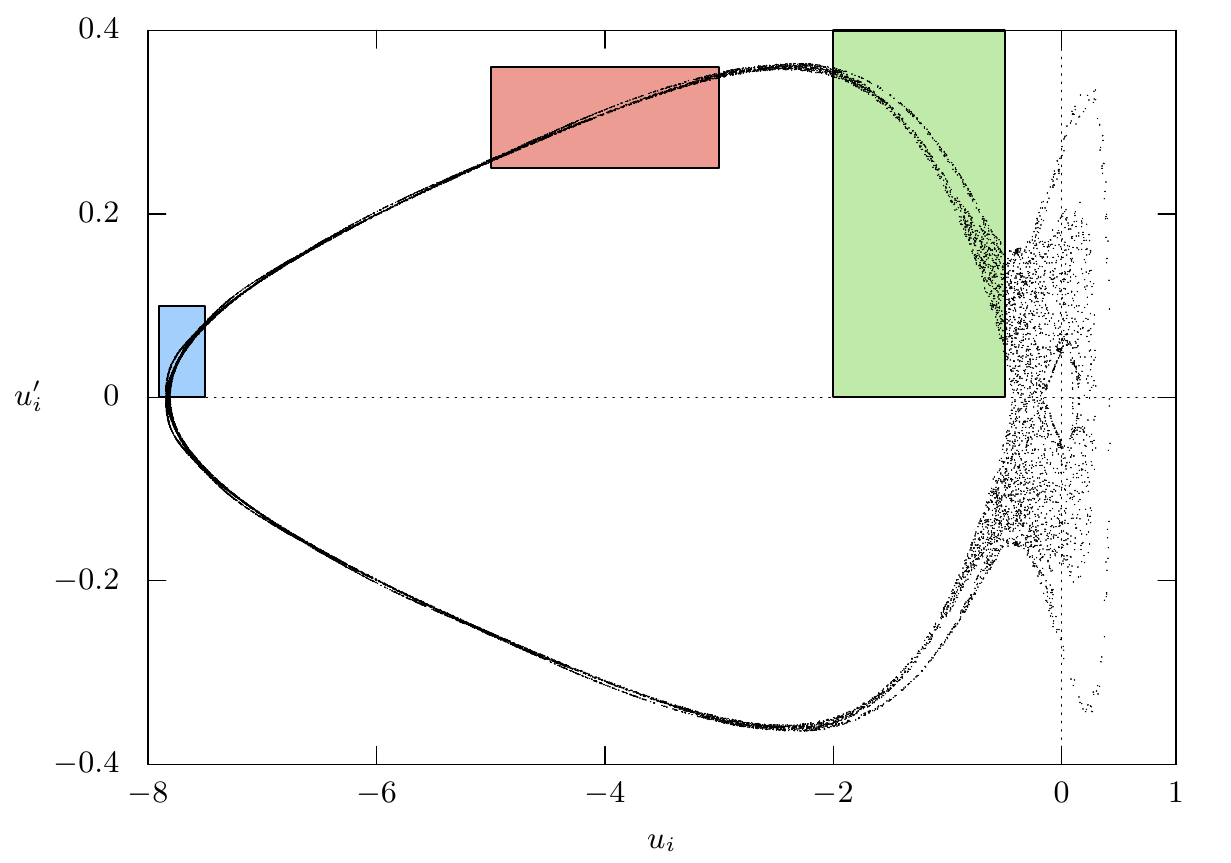}\hfill{}\includegraphics[width=0.49\textwidth]{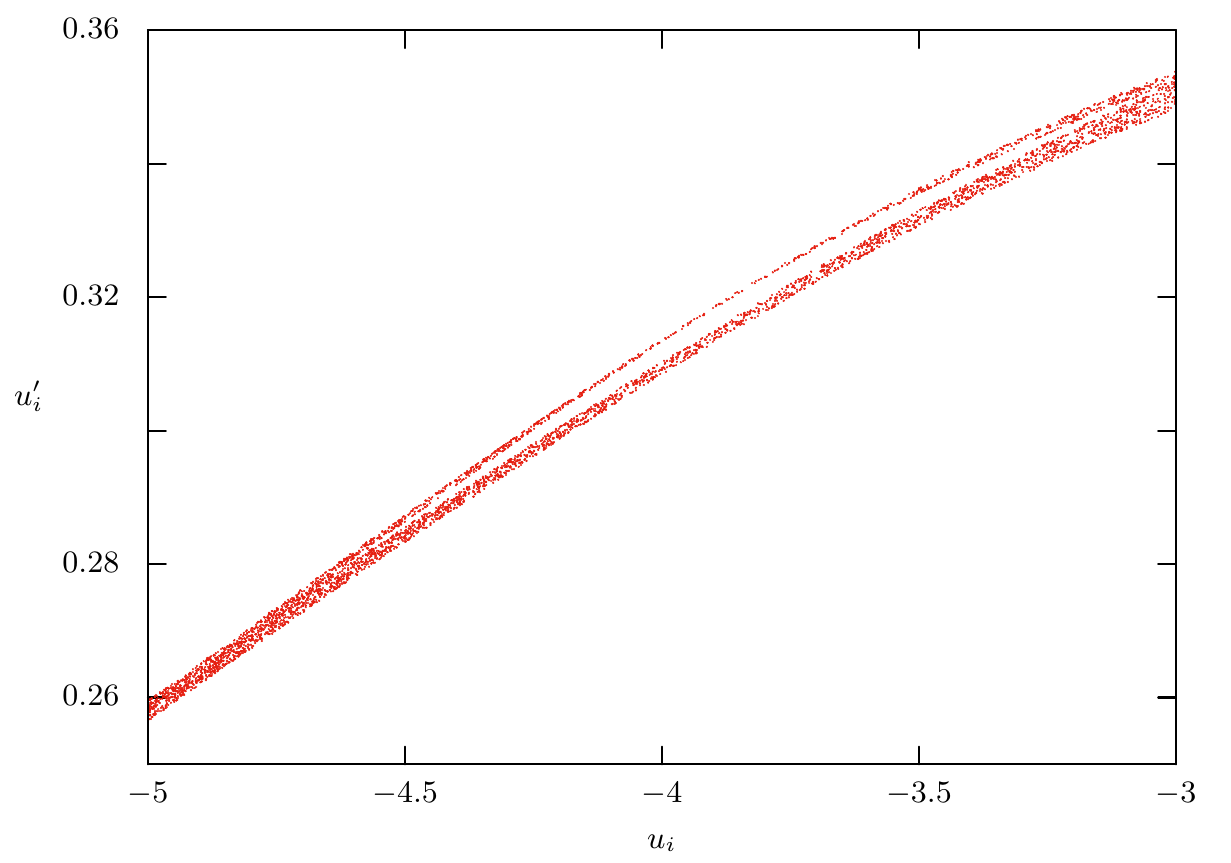} 
\par\end{centering}
\begin{centering}
\includegraphics[width=0.49\textwidth]{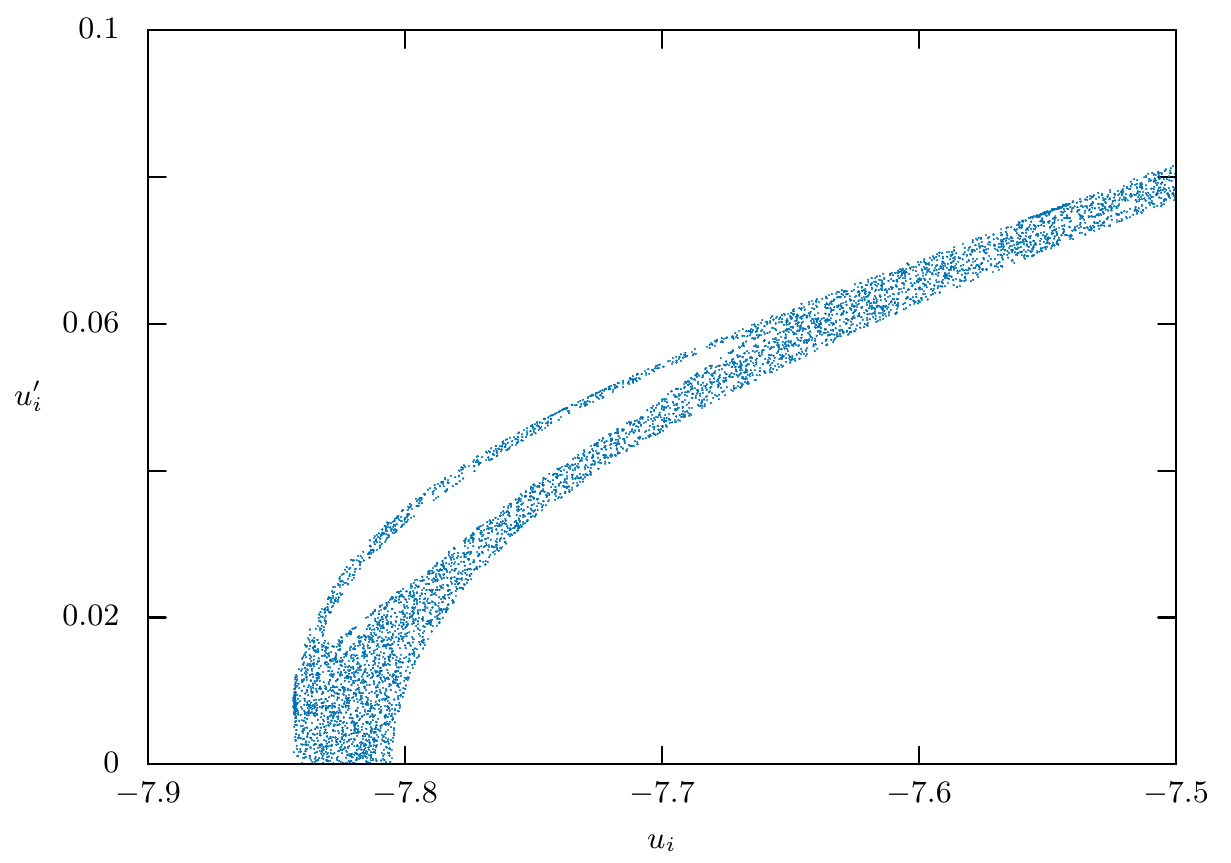}\hfill{}\includegraphics[width=0.49\textwidth]{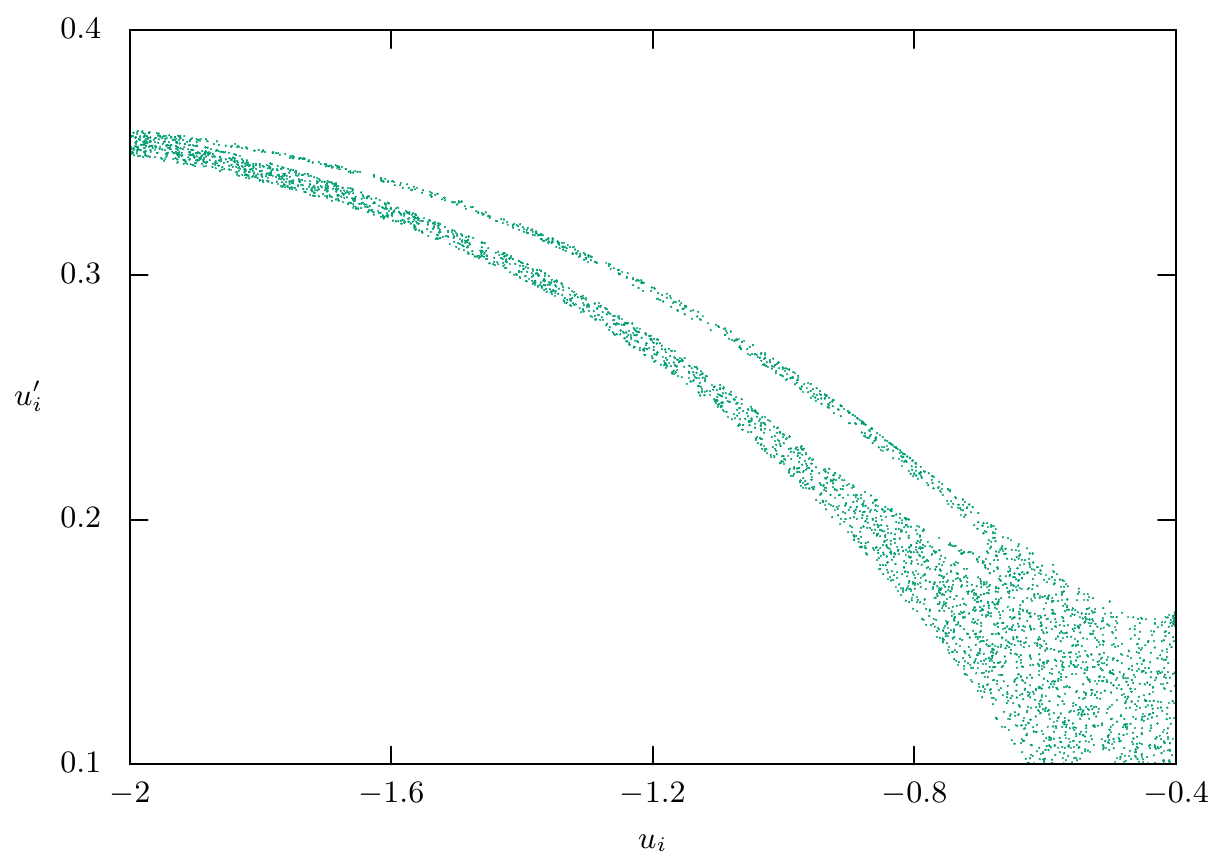} 
\par\end{centering}
\caption{\label{fig:Poincar-plots-for}Poincaré plots for the system in the
chaotic regime. Subsequent panels are zoomed out plots of the corresponding
regions shown in the first plot.}
\end{figure}

\subsection{A chaotic ion-dynamics model}

We now construct a ion-dynamical model from Eqs.(\ref{eq:cont}-\ref{eq:pois})
for the driven dust-charge case, which can exhibit chaos. We introduce
a scaled time $\tau=t-x/v_{0}$ so that the equations can be converted
to a single variable. In terms of the scaled variable, we have $\partial/\partial t\equiv\partial/\partial\tau$
and $\partial/\partial x\equiv-v_{0}^{-1}\partial/\partial\tau$.
So, Eqs.(\ref{eq:cont},\ref{eq:mom},\ref{eq:pois}) become 
\begin{eqnarray}
\frac{\partial n_{i}}{\partial\tau}-\frac{1}{v_{0}}\frac{\partial}{\partial\tau}(n_{i}u_{i}) & = & 0,\label{eq:cont-2}\\
\frac{\partial u_{i}}{\partial\tau}-\frac{u_{i}}{v_{0}}\frac{\partial u_{i}}{\partial\tau}-2\frac{\sigma}{v_{0}}\frac{\partial n_{i}}{\partial\tau} & = & \frac{1}{v_{0}}\frac{\partial\phi}{\partial\tau},\label{eq:mom-2}\\
\frac{1}{v_{0}^{2}}\frac{\partial^{2}\phi}{\partial\tau^{2}} & = & n_{e}-\delta_{i}n_{i}+\delta_{d}z_{d}(\tau),\label{eq:pois-1}
\end{eqnarray}
where for simplicity we assumed $\gamma=2$. Eq.(\ref{eq:cont-2})
can be integrated to obtain the ion density 
\begin{equation}
n_{i}=\frac{v_{0}-u_{0}}{v_{0}-u_{i}},
\end{equation}
where we have assumed that at infinity $u_{i}(\tau)\to u_{0}$ and
$n_{i}(\tau)\to n_{i0}\equiv1$ (note the normalization). Similarly
Eq.(\ref{eq:mom-2}) can be integrated to obtain the plasma potential
\begin{equation}
\phi=\frac{1}{2}\left(u_{0}^{2}-u_{i}^{2}\right)-v_{0}(u_{0}-u_{i})+2\sigma\left(1-\frac{v_{0}-u_{0}}{v_{0}-u_{i}}\right),
\end{equation}
where we have imposed the condition that at infinity (bulk plasma),
$\phi(\tau)\to0$. Using the Poisson equation and the above expressions,
we finally arrive at a coupled 2-D nonlinear differential equation
in $u_{i}$ 
\begin{equation}
u_{i}''A+u_{i}'^{2}B+C=0,\label{eq:chaos}
\end{equation}
where the `$'$' denotes $d/d\tau$ and 
\begin{eqnarray}
A & = & \frac{1}{v_{0}^{2}}\left(\mu_{i}-2\sigma\frac{\mu_{0}}{\mu_{i}^{2}}\right),\\
B & = & -\frac{1}{v_{0}^{2}}\left(1+4\sigma\frac{\mu_{0}}{\mu_{i}^{3}}\right),\\
C & = & \delta_{i}\frac{\mu_{0}}{\mu_{i}}-e^{\phi}-\delta_{d}z_{d}(\tau),
\end{eqnarray}
with $\mu_{0}=v_{0}-u_{0}$ and $\mu_{i}=v_{0}-u_{i}$. With an autonomous
term involving the time-varying dust-charge $z_{d}(\tau)$, Eq.(\ref{eq:chaos})
fulfills all the basic characteristics which are necessary for exhibition
of chaotic dynamics. The driven dust-charge term $z_{d}(\tau)$ can
be designed as 
\begin{equation}
z_{d}(\tau)=\varrho\,\cos(\nu t)^{\Delta},\quad(\Delta\in{\rm integer}),\label{eq:burst1}
\end{equation}
where $\varrho$ is the amplitude and $\nu$ is the frequency of the
driven dust-charge fluctuation. It can be numerically shown that in
absence of any dust-charge fluctuation i.e.\ when $z_{d}\equiv1$,
for $v_{0}>u_{0}$, Eq.(\ref{eq:chaos}) admits periodic solutions
(See Fig.\ref{fig:The-periodic-oscillations}).

\begin{figure}
\begin{centering}
\includegraphics[width=0.5\textwidth]{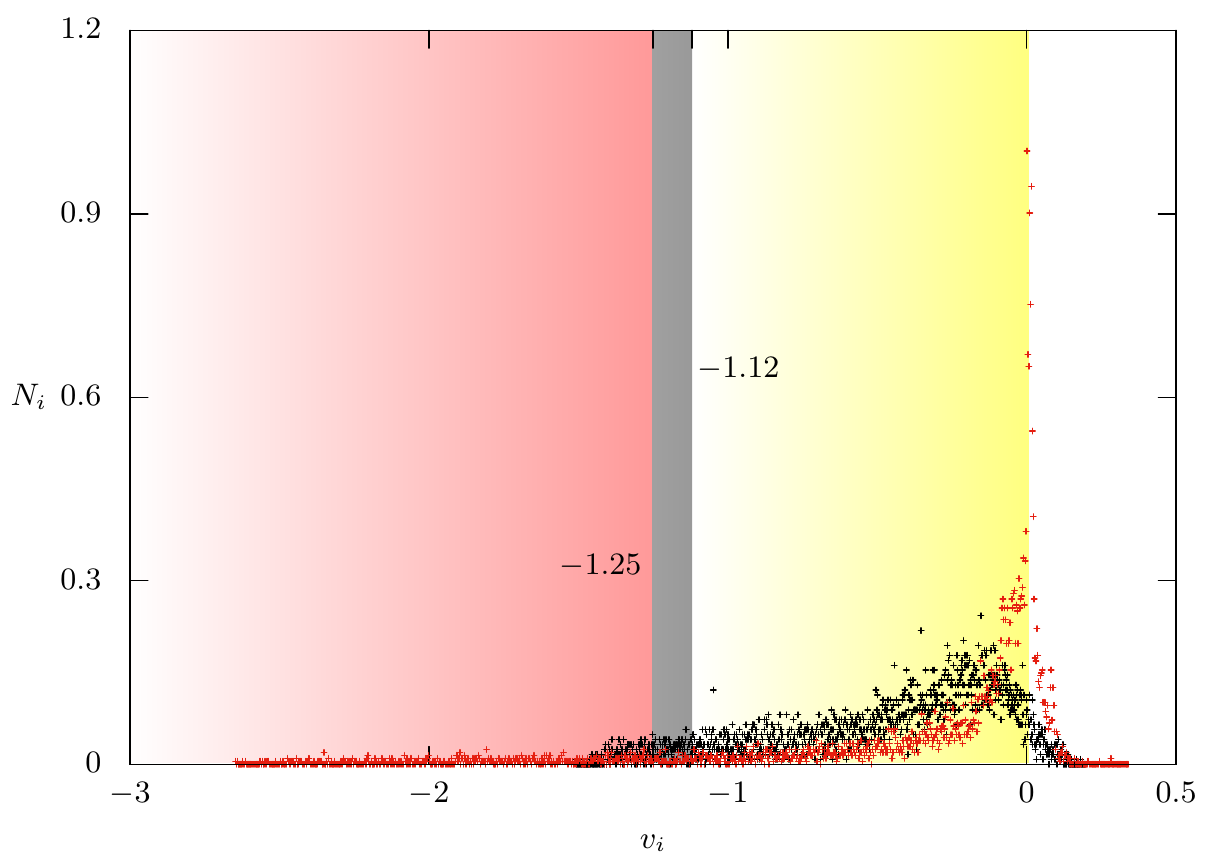}
\par\end{centering}
\caption{\label{fig:The-ion-velocity}The ion velocity distributions in the
sheath region (up to about $x\apprle10\lambda_{D}$) with and without
dust-charge fluctuation. While the black points indicate the distribution
with constant dust-charge, the red points indicate the same with dust-charge
fluctuation. Please see the accompanying text for explanation about
the different regions indicated in the figure. The sheath is on the
left $(x=0)$ and the `$-$' sign indicates left-moving ions going
toward the sheath.}
\end{figure}

\subsubsection{$v_{0}\gg u_{0}$}

We now explore the regime $v_{0}\gg u_{0}$ in presence of driven
dust-charge fluctuation. Our parameters are $\varrho\sim0.1$, which
is a small positive quantity and $\nu\sim0.4$. We take $u_{0}\sim10^{-4}$
and $v_{0}\sim1$. The parameter $\Delta$ is set to $2$. Note that
when $v_{0}\gg u_{0}$, it is as if $\tau\sim t$ and the variables
are almost constant in space. This points out to a localised disturbance
where chaotic oscillation might be observed. The rest of the plasma
parameters are as in case of the simulation.

The results of this analysis is presented in Fig.\ref{fig:The-periodic-oscillations}
and \ref{fig:The-maximal-Lyapunov}. The left panel of Fig.\ref{fig:The-periodic-oscillations},
shows the periodic oscillations when $z_{d}=1$ or equivalently when
$\varrho=0$ i.e.\ no dust-charge fluctuation. The right panel shows
the same oscillations (in black and red colors) when $\varrho=0.1$
(the other parameters are as mentioned before). One can clearly see
the sensitivity of the oscillations where both of these differ by
a factor $\sim10^{-5}$ in the initial conditions. We carry out a
Lyapunov exponent calculation \cite{strogatz-book} on Eq.(\ref{eq:chaos})
for these oscillations and the results are presented in Fig.\ref{fig:The-maximal-Lyapunov},
where the left panel shows the maximal Lyapunov exponent of the system
$l\sim0.033$, which is positive signifying chaos and in the right
panel, the phase portrait of the system is shown. In calculating the
Lyapunov exponent, we have evolved the system for $\tau\simeq2000$
with a step-size of $0.01$, gathering about $2\times10^{5}$ points.
The corresponding Poincaré plots \cite{strogatz-book} are shown in
Fig.\ref{fig:Poincar-plots-for}, which shows the typical characteristics
of a fractal construction, signifying chaos.

\section{\emph{hybrid}-PIC-MCC simulation of dusty plasma}

Our simulation model comprises of a 1-D electrostatic particle-in-cell
(PIC) code with the capabilities of having various boundary conditions
including periodic boundary \cite{suniti,suniti1,suniti2}. We note
that the usual PIC model does not have any collisional transfer of
momenta. Also the particles in a PIC model are macro-particles comprising
of a number of real-life particles. As such, any collision implemented
under the PIC formalism will actually account for collisions \emph{en
masse}. However, in a limited way, collisions can be implemented in
a PIC formalism through PIC-Monte Carlo Collision (PIC-MCC) algorithm.
It consists of using a randomized probability to account for the collisions
based on the theoretical estimation of the collision cross sections.
Multistep Monte Carlo Collision is also another way of including collisions
\cite{gatsonis94}. We rather use a hybrid method to estimate  the
collisions of dust particles with electrons and ions, which is described
as the \emph{hybrid}-PIC-MCC (\emph{h}-PIC-MCC) method \cite{suniti1,suniti2}.
The details of this algorithm and the code is described in two papers
by Changmai and Bora \cite{suniti,suniti1}.

\subsection{Dust-charge fluctuation and plasma sheath}

In order to account for the dust charge fluctuation in our simulation,
we assume that whenever a collision of the dust particle with an electron
occurs, it contributes to an increase of negative charge on the surface
of the dust particle \cite{suniti1,suniti2}. This is effected by
decreasing the number of electrons in the simulation domain accompanied
by an equivalent increase of electron dust charge number $z_{d}$.
The accumulation of positive charge on the surface of a dust particle
is, however, modeled by assuming that whenever an ion-collision occurs,
an electron is ejected from the dust particle which causes $z_{d}$
to decrease (equivalently charging the dust particle positively) and
an increase of a plasma electron in the simulation box. Hence, the
total ion number in the simulation domain remains constant while the
total electron number and dust charge fluctuate depending on the type
of collisions. Without loss of any generality, this process can be
extended up to any number of dust particles, thereby simulating an
environment of a dusty plasma. 
\begin{figure}
\begin{centering}
\includegraphics[width=0.5\textwidth]{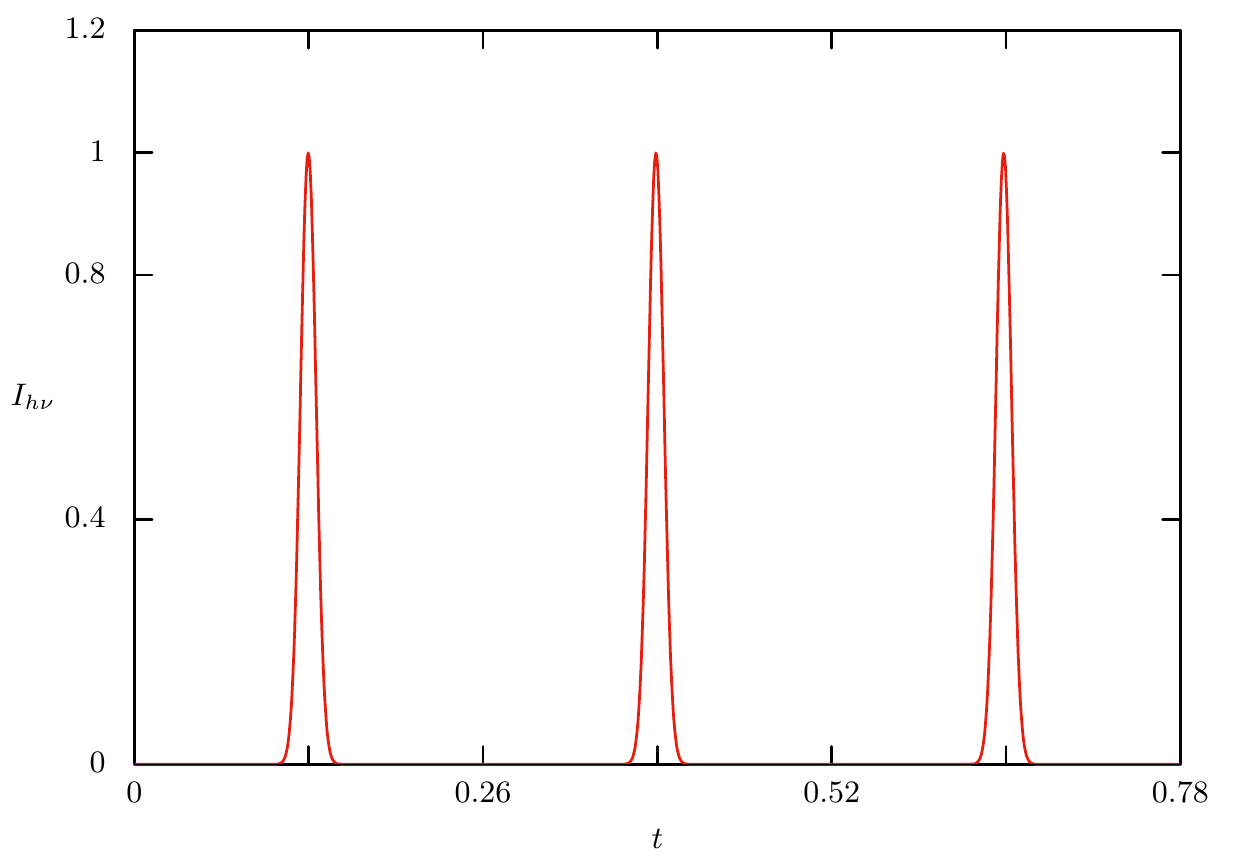}
\par\end{centering}
\caption{\label{fig:The-periodic-burst}The periodic burst of photoemission.}
\end{figure}

In our simulation, we however assume that the dust particles are cold
and stationary, which is consistent with the characteristic time scale
of the simulation i.e.\ ion-acoustic time scale and is also what
we consider in the theoretical buildup. We also introduce a randomized
probability $p_{i,e}$, which determines whether a charged particle
is absorbed by a dust particle in the event of a collision, hence
the name \emph{h}-PIC-MCC. At this point, it should be noted that
every binary collision in this formalism is actually a collision between
two macro-particles, which in reality, does not happen. Nevertheless,
in the ion acoustic time scale, this procedure is able to capture
the essential physics involving dust-ion-acoustic (DIA) dynamics.
For all practical purposes, the dust particles act as collections
of electrons, assuming that the dust particles charge to a net negative
potential as the plasma attains its equilibrium. Our prescription
for dust charging conserves the plasma quasi-neutrality condition
\begin{equation}
n_{i}=n_{e}+z_{d}n_{d}.
\end{equation}
The dust macro-particles are uniformly distributed in the domain with
zero net charge. The dust radius is fixed at $r_{d}\sim10^{-2}\lambda_{{\rm D}e}$.
As mentioned before, in Fig.\ref{fig:Charging-of-dust}, the average
charging of single dust particle is shown, calculated from the total
count of electron depletion in the simulation domain. As can be seen
from the figure, on average, a single dust particle attains an equilibrium
net charge of about $Q_{d}\sim-500e$. The equivalent dust number
density can be calculated from the quasi-neutrality condition as $n_{d}\sim4.2\times10^{12}\,{\rm m}^{-3}$. 

We now present the results of our \emph{h}-PIC-MCC simulation on the
effect of dust-charge fluctuation. We note that the Bohm criterion
for a plasma sheath with cold ion and constant dust-charge is $M>M_{c}=\sqrt{\delta_{i}}$,
which gets modified for dust-charge fluctuation \cite{Shukla}
\begin{equation}
M_{c}^{2}=\frac{-\delta_{i}+2q_{d0}+\sqrt{(3\delta_{i}-2q_{d0})^{2}+8(\delta_{i}-1)[2-(3\delta_{i}-2)/q_{d0}]}}{2[1-(\delta_{i}-1)/q_{d0}]},
\end{equation}
where $q_{d0}$ is the normalised dust-charge at the sheath edge.
For the plasma parameters as mentioned in the previous paragraph,
$M_{c}\simeq1.12$ for constant dust-charge and gets modified to $M_{c}\simeq1.25$
in presence of dust-charge fluctuation. In Fig.\ref{fig:The-ion-velocity},
we show the ion velocity distribution in the sheath region. Note that
the sheath region extends \emph{only }up to about $\sim10\lambda_{D}$
(see the first panel in Fig.\ref{fig:Plasma-potential-}). In Fig.\ref{fig:The-ion-velocity},
we show the distributions for dust particles with constant charge
(black dots) and with charge fluctuation (red dots). As expected,
the penetrating velocity of the ions into the sheath increases in
presence of dust-charge fluctuation. Different regions indicated in
the figure represent the pre-sheath region, sheath-edge for constant
dust-charge, and sheath-edge with dust-charge fluctuation. The respective
sheath-edges are defined through the values $M=M_{c}$ and the pre-sheath
region is defined when the ion velocity is just zero demarcating the
let and right-moving ions. So, in this case, we define the pre-sheath
region is where ions have \emph{just} started to move toward the left
i.e.\ toward the sheath. The number of ions in the figure is normalised
by its maximum value. 
\begin{figure}
\includegraphics[width=0.49\textwidth]{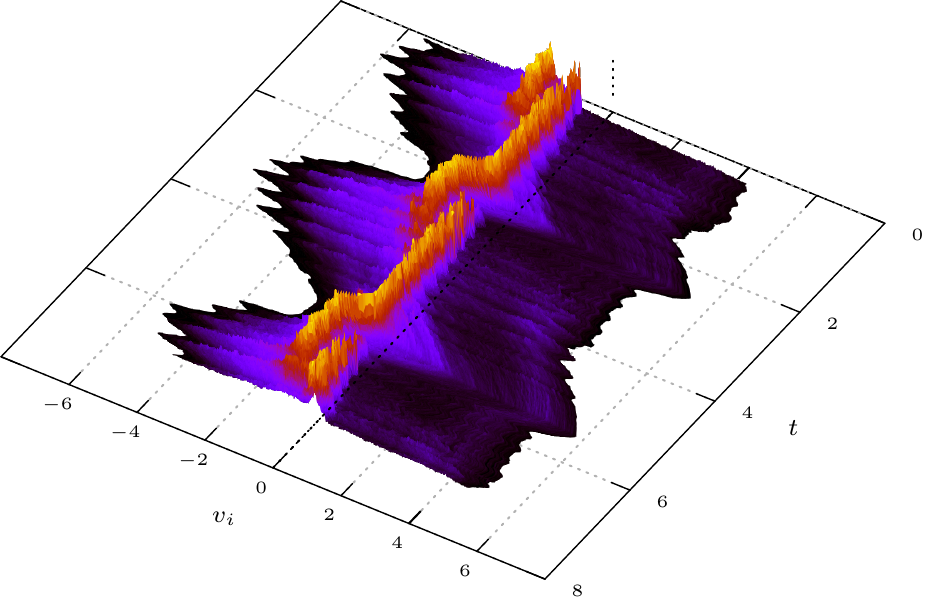}\hfill{}\includegraphics[width=0.49\textwidth]{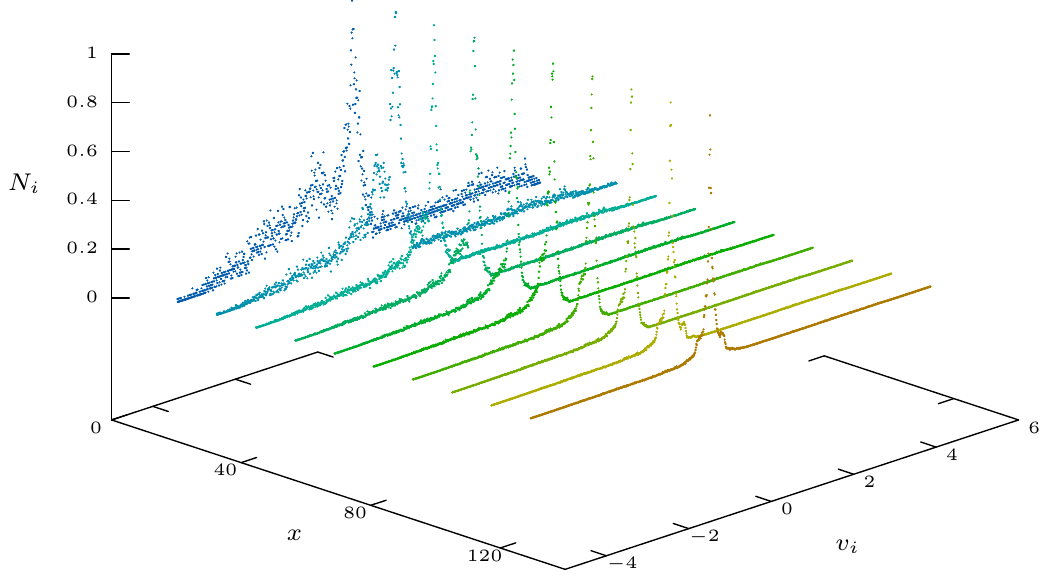}

\caption{\label{fig:The-evolution-of}The evolution of the plasma sheath in
time (first panel) and in space (second panel) in terms of ion velocity
distribution. In the first panel, the ion velocity distribution is
plotted in the near-sheath region in time, where the oscillation of
the sheath (in terms of periodic peaking of ion velocity) can be clearly
seen. In the second panel, the ion velocity distribution function
is plotted as one moves away from the sheath (the wall is at $x=0$).}
\end{figure}

\subsection{Driven dust-charge fluctuation -- Chaotic ion dynamics}

We now consider the case for driven dust-charge fluctuation, where
we drive the dust-charge fluctuation through a controlled emission
of photoelectrons from the sheath-side wall by periodically exposing
the wall to strong UV radiation. The periodic emission of photoelectrons
can be mathematically represented (in the photoemission current to
the dust particles) as 
\begin{equation}
I_{h\nu}=\varrho\,\cos(\nu t)^{2\Delta},\,\Delta\gg1\,({\rm integer}),\label{eq:burst}
\end{equation}
where $\nu$ is the periodicity of the photoemission burst (see Fig.\ref{fig:The-periodic-burst})
and $\varrho$ is its amplitude. In our case, $\nu^{-1}\sim0.02t$,
time being normalised by $\omega_{pi}^{-1}$. Eq.(\ref{eq:burst})
is same as Eq.(\ref{eq:burst1}), except that it is now in the photoemission
current to the dust particles which has the same role in our simulation
scenario as Eq.(\ref{eq:burst1}) in the theoretical formalism.
\begin{figure}
\includegraphics[width=0.49\textwidth]{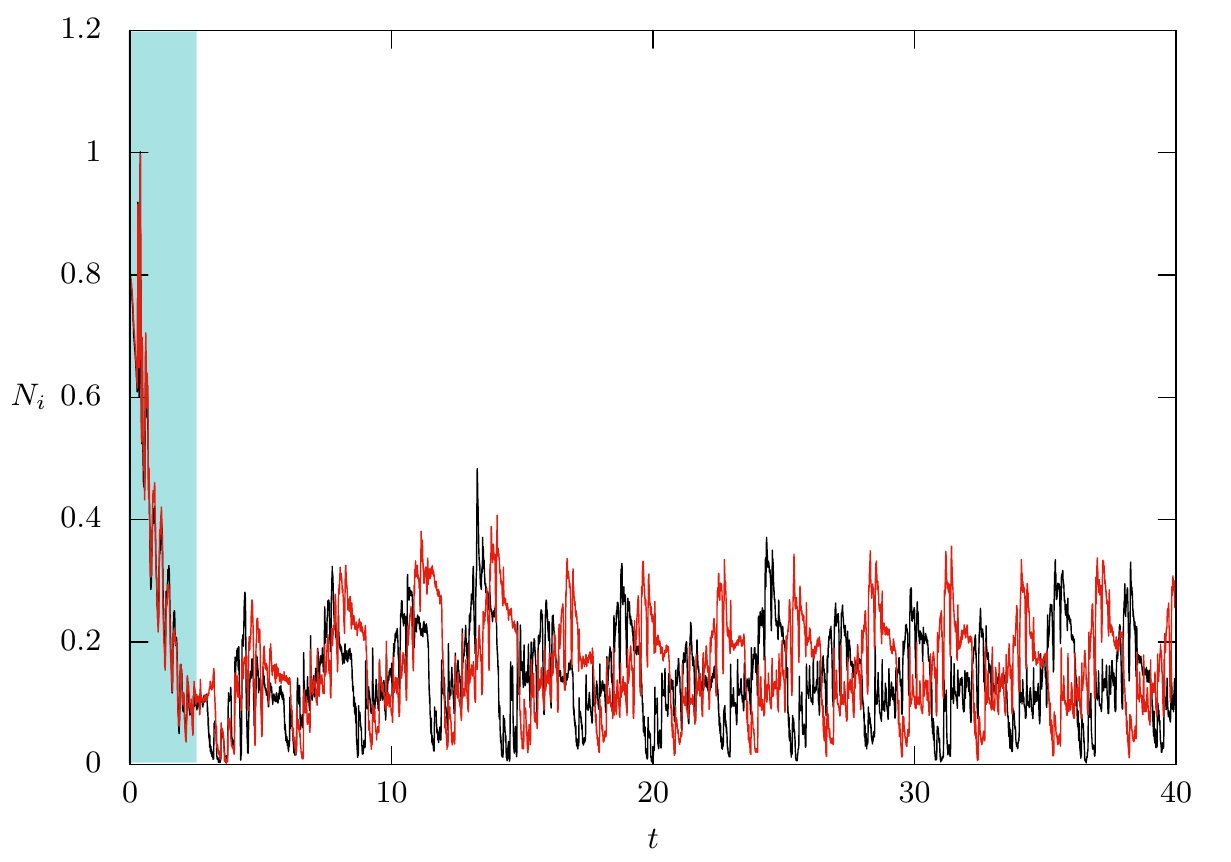}\hfill{}\includegraphics[width=0.49\textwidth]{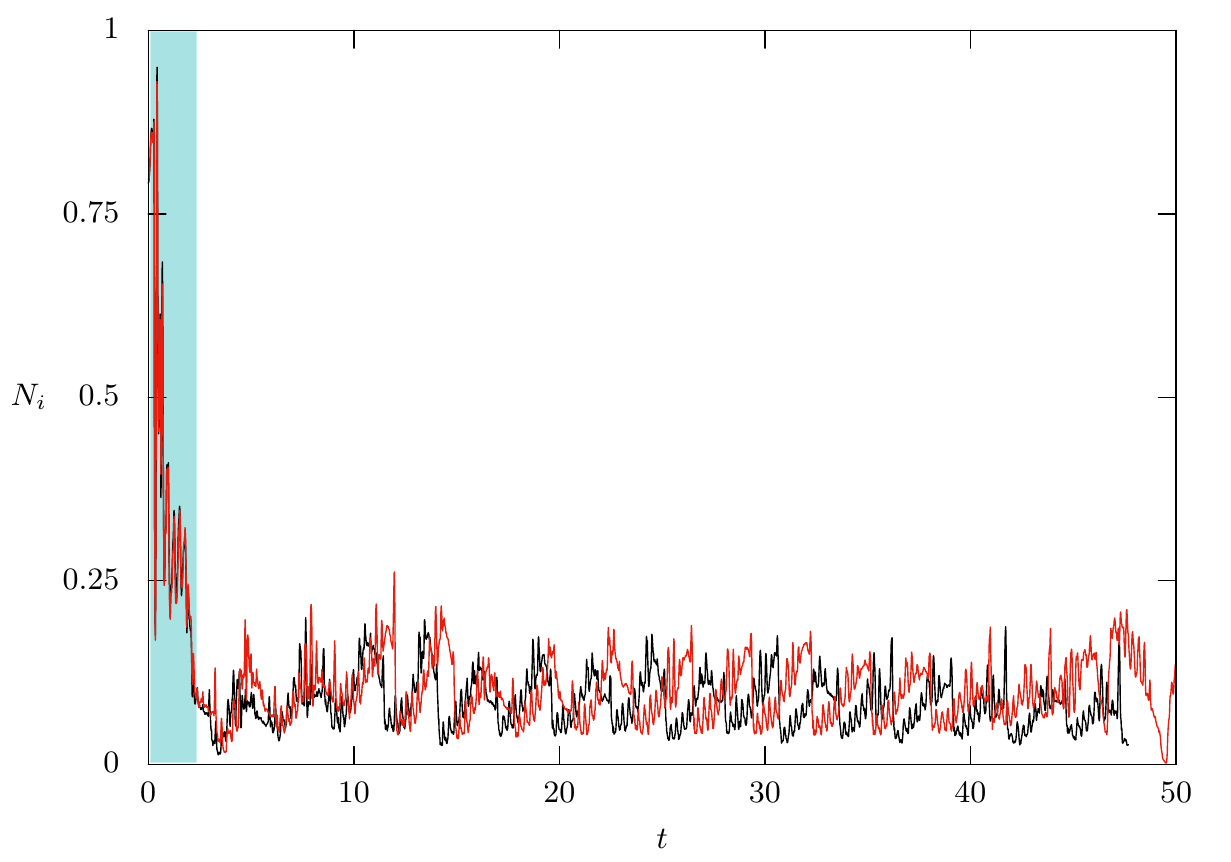}\\
 \includegraphics[width=0.49\textwidth]{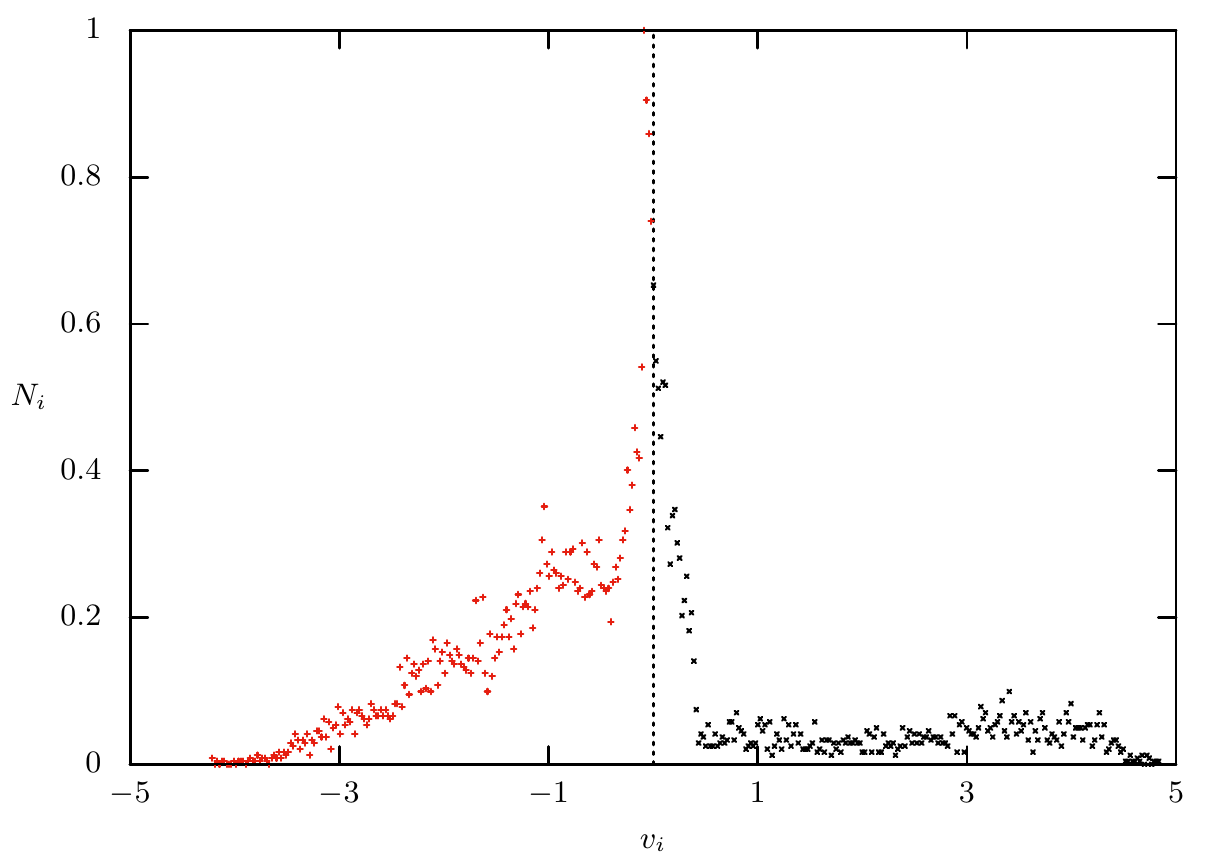}\hfill{}\includegraphics[width=0.49\textwidth]{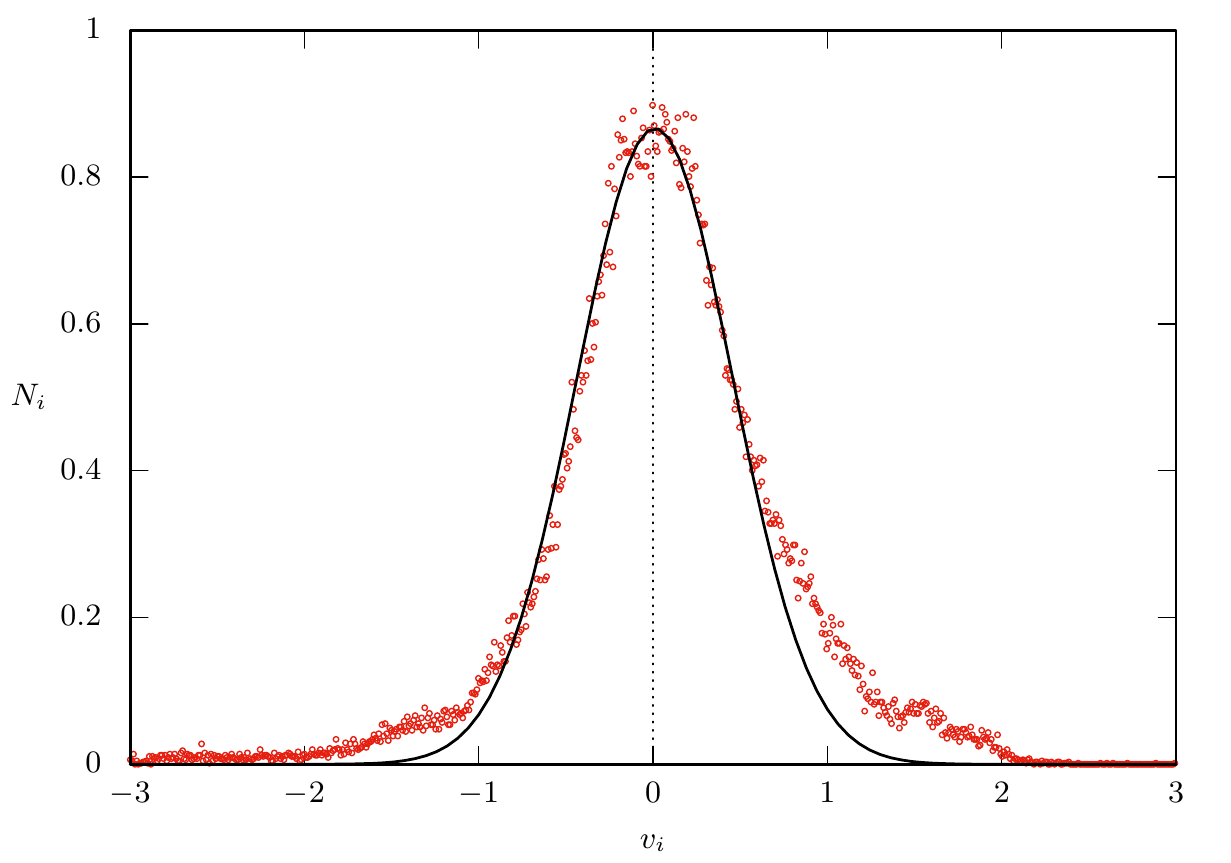}\\
 \includegraphics[width=0.49\textwidth]{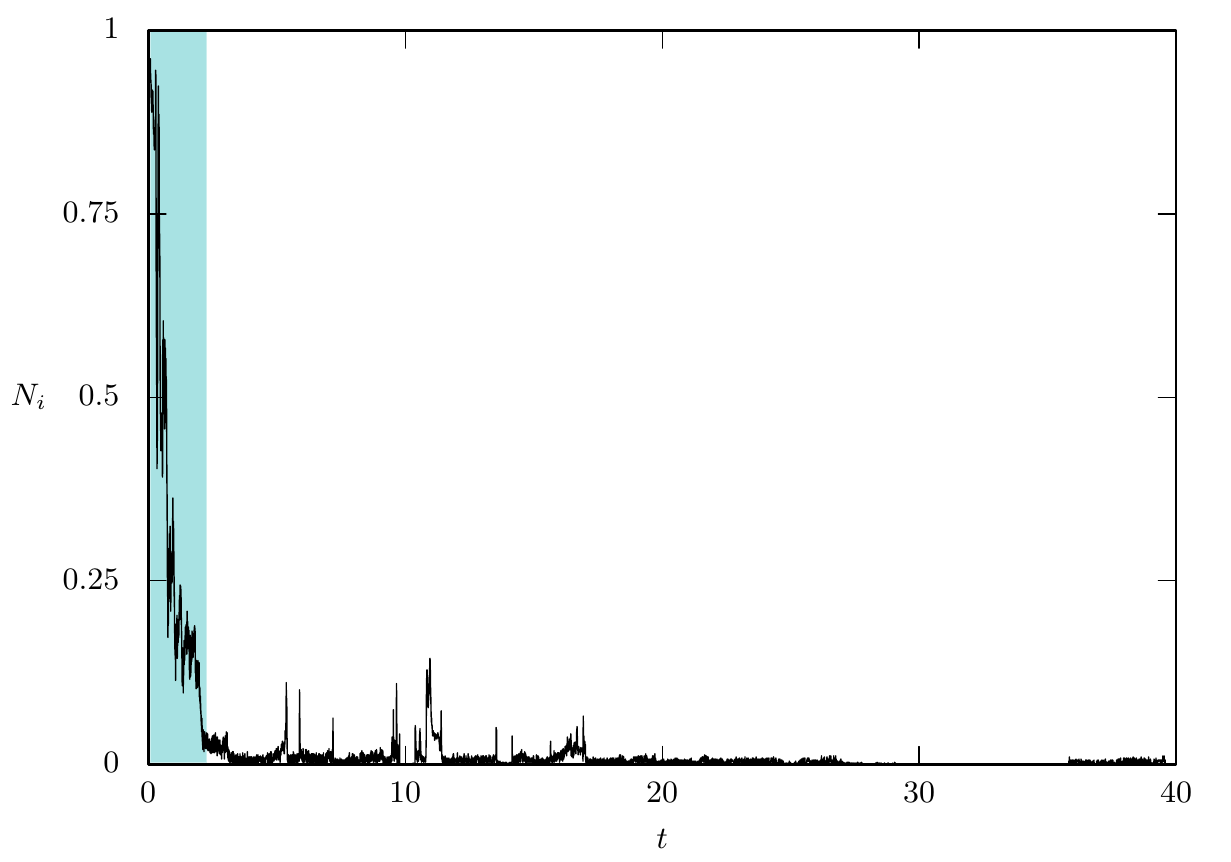}\hfill{}\includegraphics[width=0.49\textwidth]{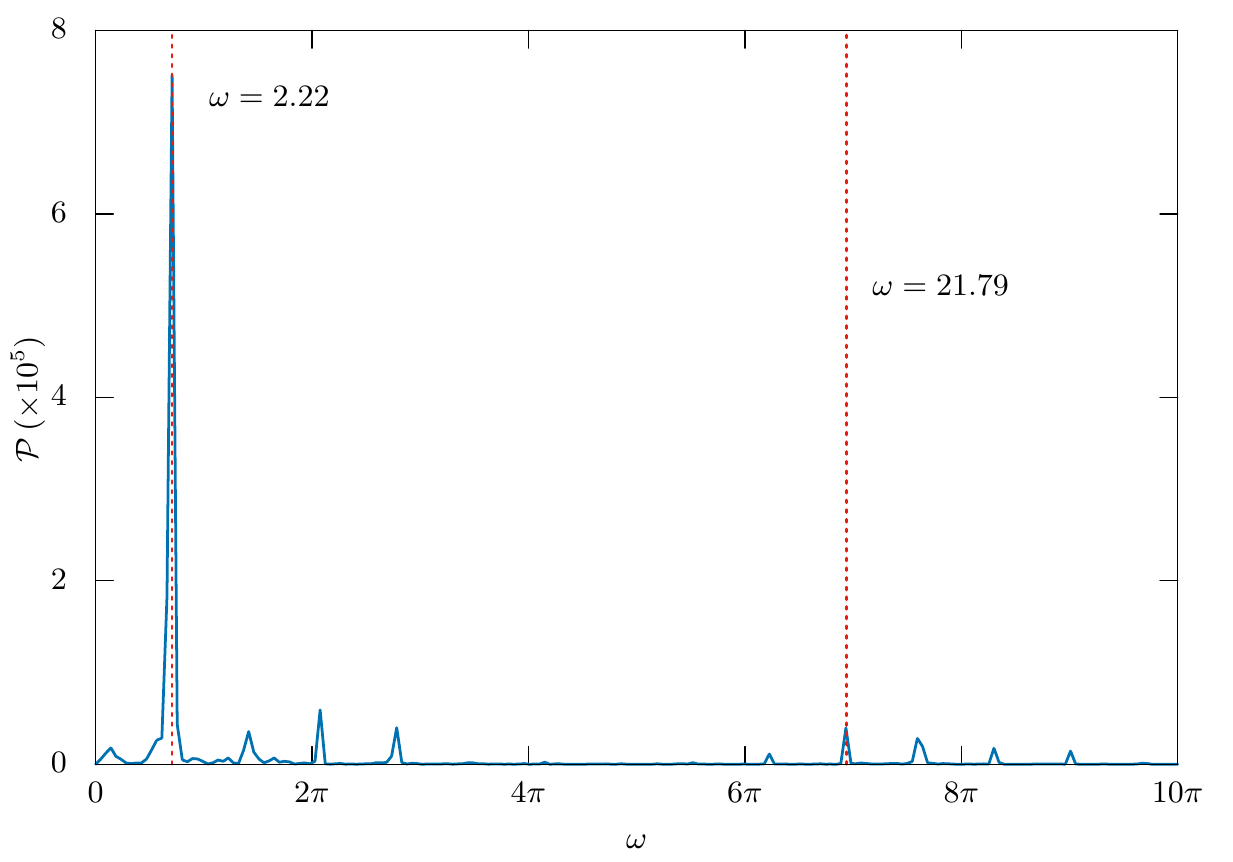}

\caption{\label{fig:Two-snapshots-of}First row: Two snapshots of the ion velocity
distribution in the sheath and the near-sheath region for two different
dust densities. The first panel is for $n_{d}\sim4.2\times10^{12}\,{\rm m}^{-3}$
and the second one is for $n_{d}\sim1.5\times10^{12}\,{\rm m}^{-3}$.
Both the plots show the characteristics signature of chaotic dynamics
as confirmed by our chaos analysis. The shaded regions in both the
plots indicate initial transient regions as the distributions evolve
in time. The change in oscillation patterns in the both the plots
corresponds to a change of $\sim10^{-4}$ in the burst duration of
the emission of photoelectrons while the periodicity of the bursts
are same in all the cases.\protect \protect \\
 ~\protect \protect \\
 Second row: The ion distribution function in the sheath region and
in the bulk plasma.\protect \protect \\
 ~\protect \protect \\
 Third row: The ion distribution function in presence of periodic
photoemission bursts as in the first cases but without any dust-charge
fluctuation. The distribution is purely stochastic (left). The right
panel shows the power spectral density ${\cal P}$ of these oscillations
shown in the left panel of the first row of this figure.}
\end{figure}

As we have periodic photoelectron bursts from the sheath-side wall,
the plasma sheath oscillates between a classical sheath and an inverse
sheath. This oscillation can be clearly seen from the reconstruction
plots of the ion velocity distribution as shown in Fig.\ref{fig:The-evolution-of}.
The first panel of the plots shows the time evolution of the ion distribution
function in the near-sheath region, while the second panel shows the
evolution of the sheath as one moves away from the wall. We can clearly
see the periodic formation of ion-velocity peaks in the first panel.
In the second panel, we can see that the ion velocity distribution
gradually approaches a Maxwellian distribution as one goes away from
the wall. In Fig.\ref{fig:Two-snapshots-of}, we show the chaotic
evolution of the ion velocity distribution in the sheath and the near-sheath
region. It should however be noted that in presence of the driven
photoemission, there is no clear sheath boundary as the sheath itself
oscillates in time. We show the power spectral density ${\cal P}$
of these oscillations (first panel, first row of Fig.\ref{fig:Two-snapshots-of})
in the second panel of the third row of Fig.\ref{fig:Two-snapshots-of}.
The spectral density shows a dominant peak at frequency $\sim0.354$
corresponding an angular frequency of $\omega\sim2.22$, which is
the large-scale oscillations that we see. The second dominant peak
in ${\cal P}$ away from the large-scale oscillations is at a frequency
$\sim3.47$ corresponding an angular frequency of $\omega\sim2.22$
which represents the fine-scale oscillations. These fine-scale oscillations
actually correspond to the frequency of the driven photoemission bursts
frequency shown in Fig.\ref{fig:The-periodic-burst}, represented
by Eq.(\ref{eq:burst}). So, what we see is that the periodic bursts
of photoemission of electrons are exciting a very low frequency large-scale
oscillations in ione velocity through dust-charge fluctuations.

\begin{figure}
\begin{centering}
\includegraphics[width=0.5\textwidth]{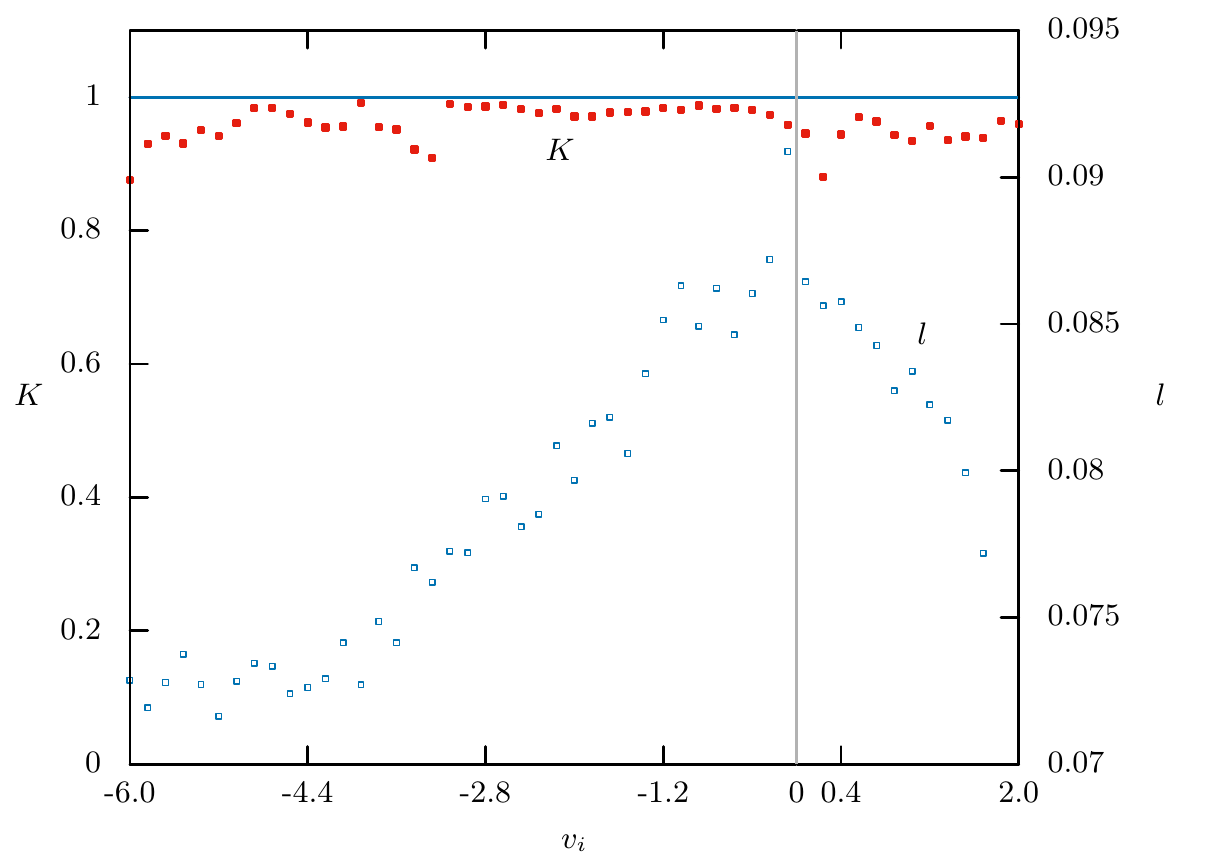}
\par\end{centering}
\caption{\label{fig:An-estimate-of}An estimate of the dominant Lyapunov exponent
as determined through Wolf's algorithm (right axis) and the $K$ values
from the 0-1 test (left axis) for the two sampled data shown in the
first row of Fig.\ref{fig:Two-snapshots-of}.}
\end{figure}

We initially perform two different tests for chaos on the ion velocity
distributions -- (\emph{a}) estimation of the dominant Lyapunov exponent
$l$ from a time series with the help of Wolf's algorithm \cite{wolf}
and (\emph{b}) perform 0-1 test for chaos. While Wolf's method calculates
the dominant Lyapunov exponent through reconstruction of the phase
space (also referred to as \emph{delay reconstruction}) from the time
series data and detection of orbital divergence, the 0-1 test estimates
the growth rate of divergence from the time-averaged mean square displacement
of the time series data. We use Wolf's algorithm as provided by the
authors through their very well-developed Matlab / Octave interface.
In case of the 0-1 test, we use a very recent work involving Chaos
Decision Tree Algorithm by Toker et al. \cite{toker} to rule out
stochasticity and random noise. Briefly, in 0-1 test, two 2-D system
$p(n)$ and $q(n)$ are derived from the 1-D time series data $v(n)$
for $n=1,2,\dots$ 
\begin{eqnarray}
p(n+1) & = & p(n)+v(n)\,\cos(cn),\label{eq:ktest1}\\
q(n+1) & = & q(n)+v(n)\,\sin(cn),\label{eq:ktest2}
\end{eqnarray}
where $c\in[0,2\pi]$ is random. For a particular $c$, the solution
to Eqs.(\ref{eq:ktest1},\ref{eq:ktest2}) requires 
\begin{eqnarray}
p_{c}(n) & = & \sum_{j=1}^{n}v(j)\,\cos(jc),\\
q_{c}(n) & = & \sum_{j=1}^{n}v(j)\,\sin(jc).
\end{eqnarray}
It can be shown that if $v(n)$ are regular, $p,q$ are bounded, while
they display asymptotic Brownian motion if $v(n)$ are chaotic. The
time-averaged mean squared displacement of $p,q$ is then calculated
as 
\begin{equation}
{\cal M}_{c}(n)=\frac{1}{N}\sum_{j=1}^{N}\left([p_{c}(j+n)-p_{c}(j)]^{2}+[q_{c}(j+n)-q_{c}(j)]^{2}\right)+\sigma\eta_{n},
\end{equation}
where $\eta_{n}\in[-1/2,+1/2]$ is a uniform random deviate and $\sigma$
is the noise level for a total of $N$ number of sampled data in the
time series. Finally the growth rate $K$ is calculated as 
\begin{equation}
K=\lim_{n\to\infty}\frac{\log{\cal M}_{c}(n)}{\log n}.
\end{equation}
For chaotic data the median $K\to1$ and for periodic system $K\to0$. 

In Fig.\ref{fig:An-estimate-of}, we show the determined values of
$l$ and $K$ from these two tests and as can be seen, both tests
points to the fact that the sampled data are indeed chaotic in nature. 

\subsubsection{Recurrence plots}

As a final test, we construct the recurrence plots \cite{eckmann,marwan}
for the oscillations shown in Fig.\ref{fig:Two-snapshots-of}. The
purpose of a recurrence plot is to visualize the recurrences of a
dynamical system. It is a very powerful tool which enables us to construct
complex dynamical pattern from a single time series. In summary, a
recurrence plot is based on the following recurrence relation \cite{marwan}
\begin{equation}
\bm{R}_{ij}=\begin{cases}
1: & \vec{x}_{i}\approx\vec{x}_{j},\\
0: & \vec{x}_{i}\not\approx\vec{x}_{j}
\end{cases}\quad i,j=1,2,\dots,N,
\end{equation}
where the $\left\{ \vec{x}_{i}\right\} _{i=1}^{N}$ is a system in
its phase space and $N$ is the number of considered states. Essentially
$\bm{R}_{i,j}$ is a Heaviside function which depends on a threshold
condition $\varepsilon$ which determines whether $\vec{x}_{i}\approx\vec{x}_{j}$.
If we assume that the state of a dynamical system $\vec{x}(t)$ is
specified by $d$ components, we can a form vector with these components
\cite{marwan}
\begin{equation}
\vec{x}(t)=(x_{i}(t))^{T},\quad i=1,2,\dots,d
\end{equation}
in the $d$-dimensional phase space. In a purely mathematically constructible
setup, all these components are known and one can easily construct
the phase space. However, in experiments and in a simulation like
ours, we have only \emph{one} time series with only one observable,
which in this case is the ion velocity. So, we have one discrete time
series 
\begin{equation}
u_{i},\quad i=1,2,\dots,N,
\end{equation}
with $\Delta t$ as the sampling interval on the basis of which we
have to reconstruct the phase space with the help of a time delay,
as mentioned before \cite{marwan}
\begin{equation}
\vec{x}_{i}=\sum_{j=1}^{m}u_{i+(j-1)\tau}\vec{e}_{j},\quad\vec{e}_{i}\cdot\vec{e}_{j}=\delta_{ij},
\end{equation}
where $m$ is the embedding dimension, $\tau$ is the time delay,
and $\vec{e}_{j}$ are the unit vectors which span an orthogonal system.
For $m\geqslant2D_{2}+1$, where $D_{2}$ is the correlation dimension
of the underlying attractor, with the help of Taken's theorem one
can show the existence of a \emph{diffeomorphism} between the original
and the reconstructed phase space or simply speaking, one can use
the reconstructed attractor to study the original one but in a different
coordinate system. So, our recurrence plot is then defined by the
relation 
\begin{equation}
\bm{R}_{ij}(\varepsilon)=H\left(\varepsilon-\left\Vert \vec{x}_{i}-\vec{x}_{j}\right\Vert \right),\quad i,j=1,2,\dots,N,
\end{equation}
where $N$ is the number of measured points , $H(x)$ is the Heaviside
step function, and $\left\Vert \cdot\right\Vert $ denotes norm. In
our case, we have used an Euclidean $L_{2}$ norm.

\begin{figure}[H]
\includegraphics[width=0.49\textwidth]{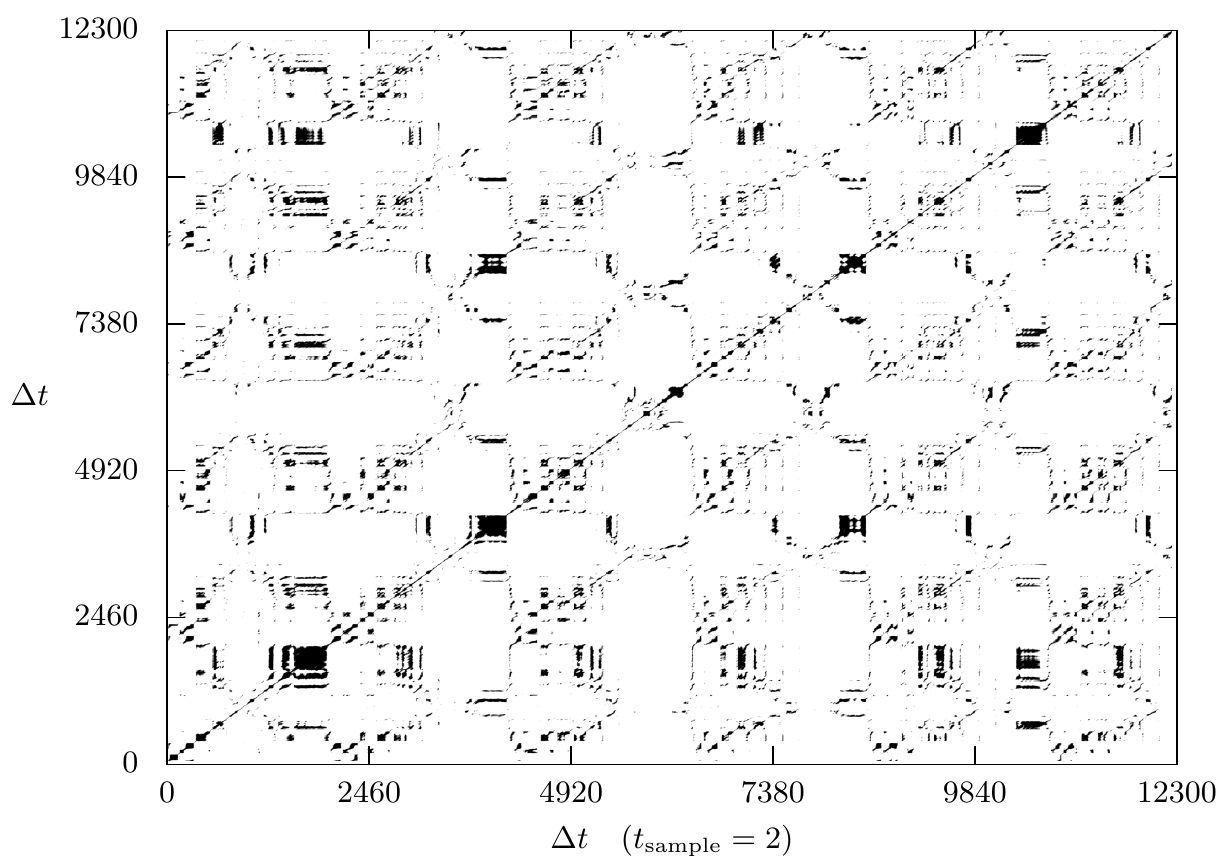}\hfill{}\includegraphics[width=0.49\textwidth]{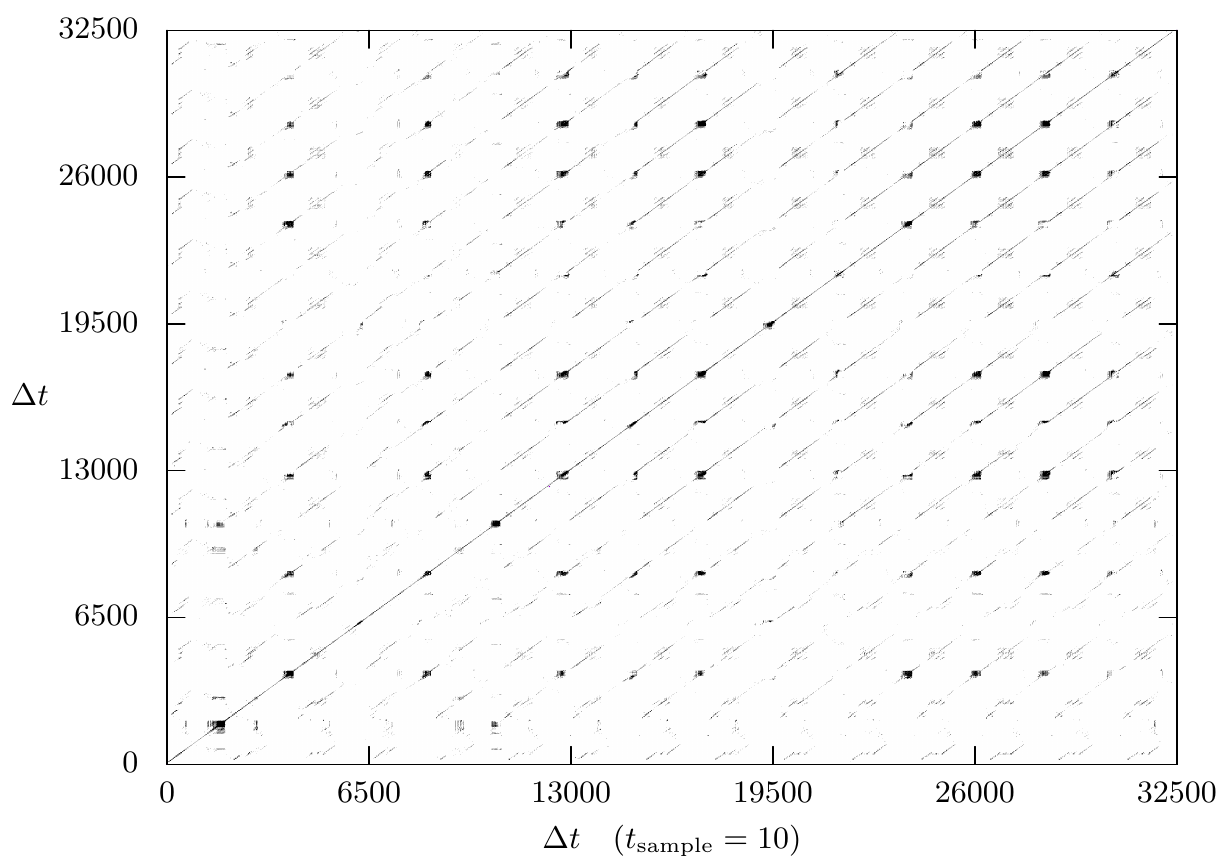}\\
 \includegraphics[width=0.49\textwidth]{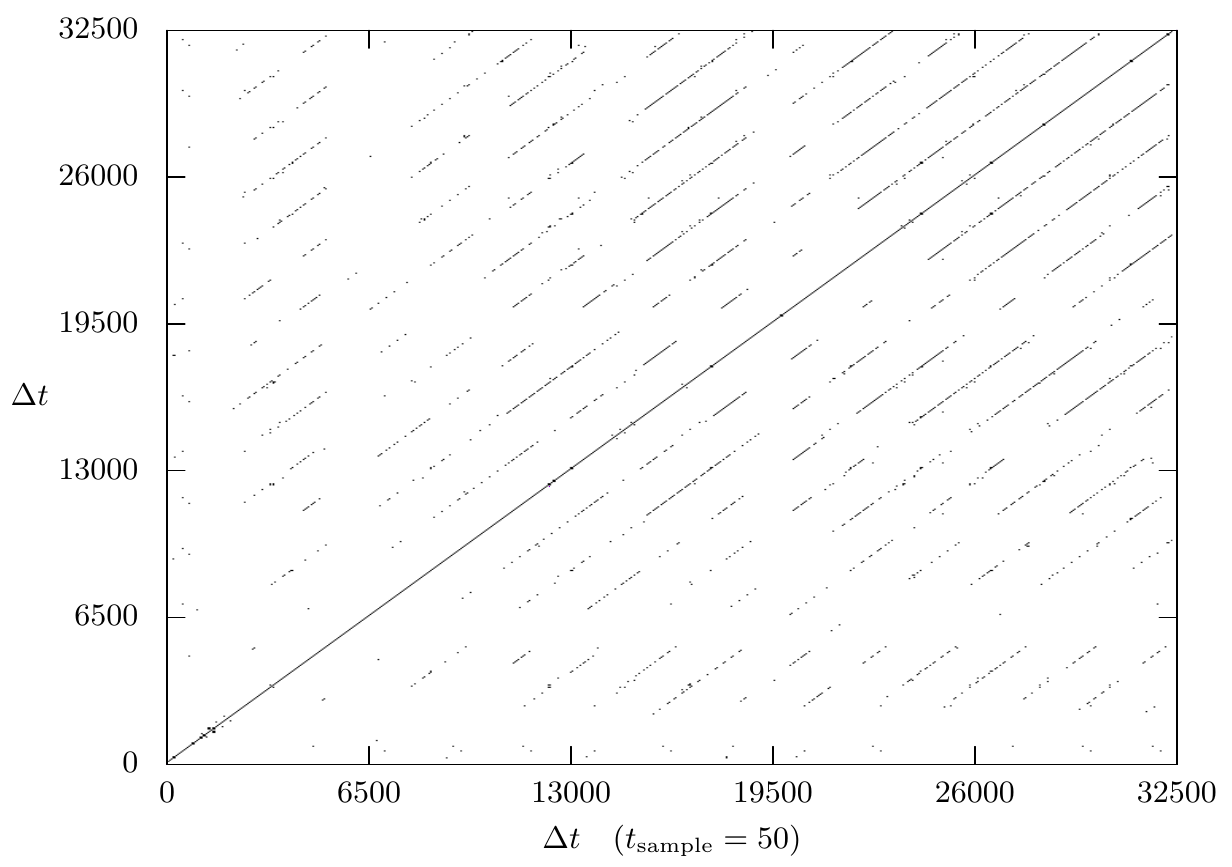}\hfill{}\includegraphics[width=0.49\textwidth]{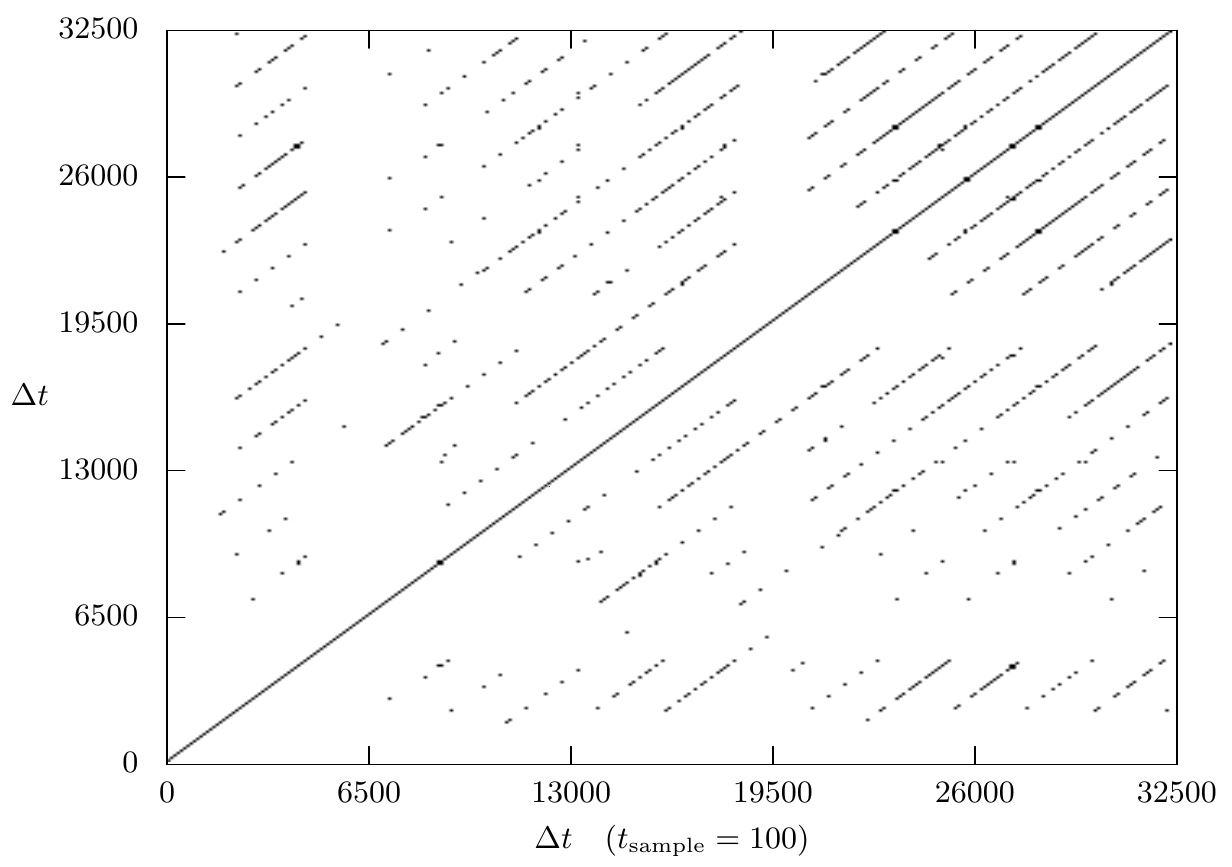}\\
 \includegraphics[width=0.49\textwidth]{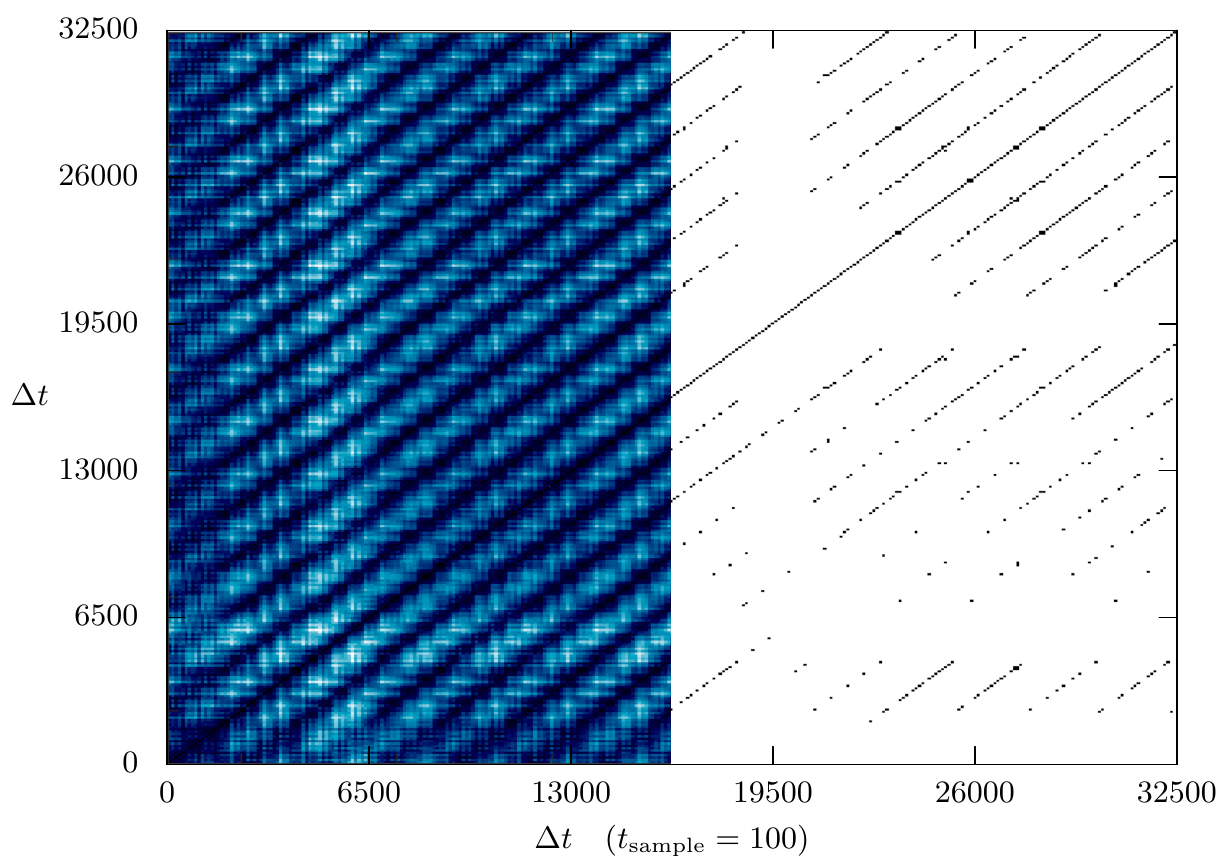}\hfill{}\includegraphics[width=0.49\textwidth]{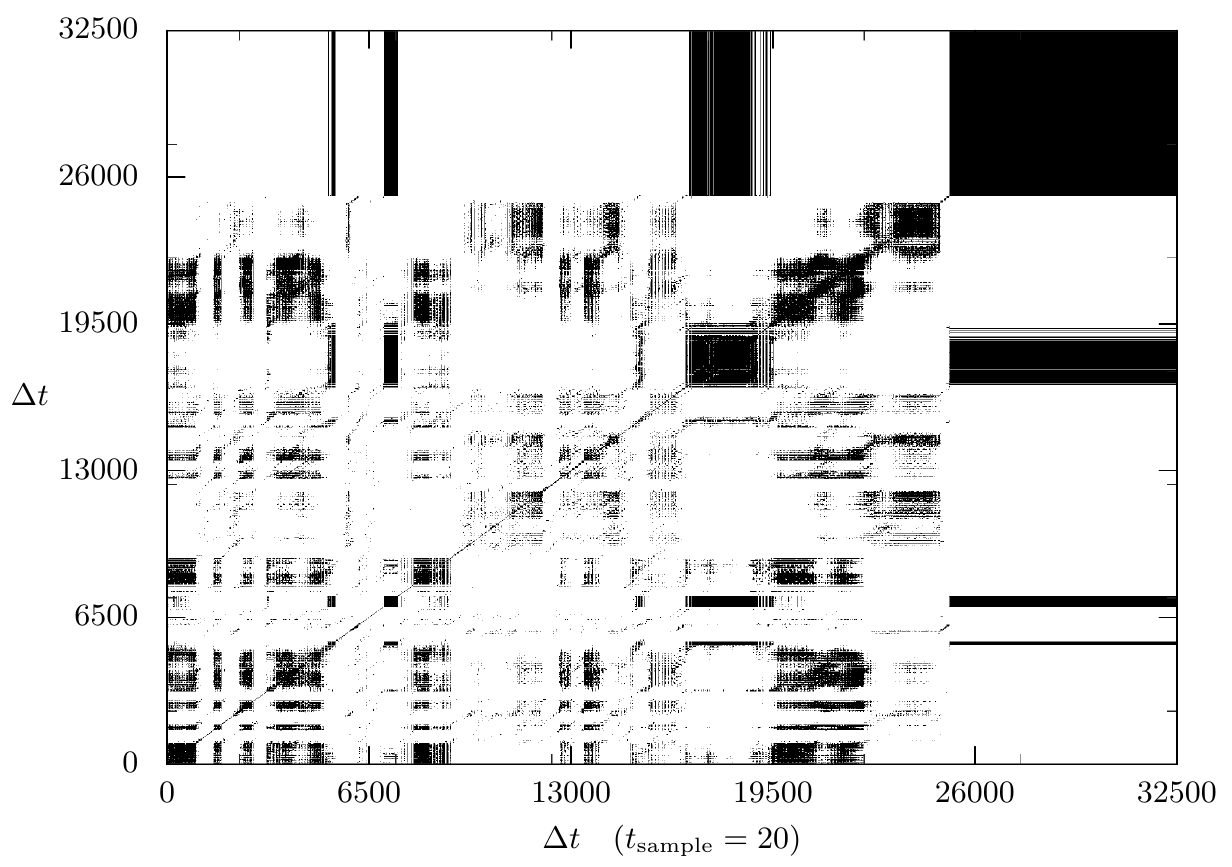}

\caption{\label{fig:First-row:-Two}First \& second rows: Recurrence plots
for the chaotic oscillation shown in Fig.\ref{fig:Two-snapshots-of}
for different sampling intervals.\protect \protect \\
 ~\protect \protect \\
 Third row: Same as the second plot of the second row, superimposed
with a map with no threshold, displaying all recurrences (left) and
a plot for a non-chaotic oscillation with constant dust-charge, displaying
the signature for Brownian motion-like pattern.}
\end{figure}

In Fig.\ref{fig:First-row:-Two}, we have shown 6 recurrence plots
for the chaotic oscillations shown in Fig.\ref{fig:Two-snapshots-of}.
The embedding dimension is chosen to be $4$ and the delay used is
$5$. The threshold is used about $\sim8.9\%$ of the maximum amplitude
of the oscillations. Clockwise from the top in the figure, we have
used a sampling interval $t_{{\rm smaple}}=2,10,50,100$ respectively
over a total of $32500$ time steps $\Delta t$. We can see that as
the sampling interval increases, the plots increasingly reveal the
signature of a chaotic oscillation, signifying the small fine-scale
periodic oscillations in the first plot with geometric recurrences
when the sampling interval is small (the first plot in the first row)
and a fractal pattern in the last of these four plots (the second
plot in the second row). The first plot in the third row is similar
to the second plot of the second row except that we have superimposed
the corresponding plot with a map having no threshold (all recurrences).
The last plot in the figure represents the one for ion velocity distribution
time series with constant dust charge, corresponding to the last plot
of Fig.\ref{fig:Two-snapshots-of}. As one can see that this plot
is similar to the one with random dynamics similar to Brownian motion
signifying stochastic behaviour \cite{marwan}.

\section{Conclusion}

To summarise, in this work in brief, we consider a plasma model where
electrons and ions are both thermal particles having cold and stationary
dust grains with constant dust density. The time scale of our interest
is ion-acoustic time scale. We develop the sheath equations and solve
them numerically with a hybrid approach coupling the initial value
problem of dust-charging equation with the boundary value problem
of nonlinear Poisson equation. The whole theoretical formalism has
been also replicated using a \emph{hybrid}-PIC-MCC simulation and
the theoretical results are found to be in good agreement with the
simulation results.

Next, we consider that case for \emph{driven }dust-charge fluctuation,
which is the primary focus of this work. We use the plasma sheath
as a candidate to induce a periodic dust-charge fluctuation through
photoemission. The photoemission occurs when the sheath-side wall
is exposed to UV radiation. So, by externally exposing the wall to
UV radiation, the dust-charge fluctuation can be driven at an external
frequency. Here, in the ion-acoustic regime, we focus on the ion dynamics
in the sheath and the pre-sheath regions. We show that the relatively
high frequency, of the order of about $\sim3.5\omega_{pi}$, bursts
of photoemission of electrons from the plasma sheath excite a chaotic
super-harmonic large low-frequency wave in ion velocity through dust-charge
fluctuation, which is confined to the sheath and the pre-sheath regions.
With the help of appropriate analysis, we have established that the
oscillations are indeed chaotic which owes its existence to the harmonic
variation of the dust-charge fluctuation. The oscillations remains
stochastic for constant dust-charge showing the in absence of sut-charge
fluctuation, the sheath oscillations can not effectively propagate
to the plasma.

\end{document}